\documentclass{emulateapj}
\usepackage{amssymb, amsmath, epsfig, rotating}

\usepackage{color}
 \def\red#1 {\textcolor{blue}{#1}~}

\newbox\grsign \setbox\grsign=\hbox{$>$} \newdimen\grdimen \grdimen=\ht\grsign
\newbox\simlessbox \newbox\simgreatbox \newbox\simpropbox
\setbox\simgreatbox=\hbox{\raise.5ex\hbox{$>$}\llap
     {\lower.5ex\hbox{$\sim$}}}\ht1=\grdimen\dp1=0pt
\setbox\simlessbox=\hbox{\raise.5ex\hbox{$<$}\llap
     {\lower.5ex\hbox{$\sim$}}}\ht2=\grdimen\dp2=0pt
\setbox\simpropbox=\hbox{\raise.5ex\hbox{$\propto$}\llap
     {\lower.5ex\hbox{$\sim$}}}\ht2=\grdimen\dp2=0pt
\def\simgreat{\mathrel{\copy\simgreatbox}}
\def\simless{\mathrel{\copy\simlessbox}}

\def\apj{ApJ}

\def\apjs{ApJS}
\def\aj{AJ}
\def\nat{Nature}
\def\mnras{MNRAS}

\def\aap{{A\&A}}
\def\prd{PRD}

\begin{document}

\title{Foreground Model and Antenna Calibration Errors in the Measurement of the Sky-Averaged $\lambda21$\,cm Signal at z$\sim$20}
\author{G.~Bernardi\altaffilmark{1,2,3}, M.~McQuinn\altaffilmark{4,5} \& L.J.~Greenhill\altaffilmark{3}}

\altaffiltext{1} {SKA SA, 3rd Floor, The Park, Park Road, Pinelands, 7405, South Africa; gbernardi@ska.ac.za}
\altaffiltext{2} {Department of Physics and Electronics, Rhodes University, PO Box 94, Grahamstown, 6140, South Africa}
\altaffiltext{3} {Harvard-Smithsonian Center for Astrophysics, 60 Garden Street, Cambridge, MA, 02138, USA}
\altaffiltext{4} {Einstein Fellow}
\altaffiltext{5} {Department of Astronomy, University of California, Berkeley, CA 94720}

\begin{abstract} 
The most promising near-term observable of the cosmic dark age prior to widespread reionization ($z\sim15-200$) is the sky-averaged $\lambda21$\,cm background arising from hydrogen in the intergalactic medium.  Though an individual antenna could in principle detect the line signature, data analysis must separate foregrounds that are orders of magnitude brighter than the $\lambda21$\,cm background (but that are anticipated to vary monotonically and gradually with frequency, e.g., they are considered ``spectrally smooth").  Using more physically motivated models for foregrounds than in previous studies, we show that the intrinsic spectral smoothness of the foregrounds is likely not a concern, and that data analysis for an ideal antenna should be able to detect the $\lambda21$\,cm  signal after subtracting a $\sim 5^{\rm th}$ order polynomial in $\log \nu$. However, we find that the foreground signal is corrupted by the angular and frequency-dependent response of a real antenna.   The frequency dependence complicates modeling of foregrounds commonly based on the assumption of spectral smoothness.  Our calculations focus on the Large-aperture Experiment to detect the Dark Age (LEDA), which combines both radiometric and interferometric measurements. We show that statistical uncertainty remaining after fitting antenna gain patterns to interferometric measurements is not anticipated to compromise extraction of the $\lambda 21$\,cm signal for a range of cosmological models after fitting a $7^{\rm th}$ order polynomial to radiometric data.  Our results generalize to most efforts to measure the sky-averaged spectrum.
\end{abstract}


\section{introduction}

The predicted transition from the cosmological dark age ($z\simgreat 30$) to the epoch of reionization (EoR; $z \simless 15$) was marked by the appearance of the first generation of stars, supernovae, and black holes.  These objects initiated a reheating and reionization of the intergalactic medium (IGM; e.g., \citealt{madau97}). The $\lambda$21\,cm transition of hydrogen is potentially sensitive to these processes even at $z\simgreat 15$, redshifts that likely cannot be probed with other known observables. Most theoretical studies have focused on the origin of (and detection prospects for) angular fluctuations in $\lambda$21\,cm brightness (\citealt{madau97,zaldarriaga04,mcquinn06,furlanetto06,morales10}). However, the sky-averaged spectrum of $\lambda$21\,cm brightness encodes independent information  \citep{shaver99,gnedin04,sethi05,furlanetto06a,pritchard10,mirocha13}.  

In fact, the instrumentation required to detect the sky-averaged signal differs markedly from that needed to detect spatial fluctuations in the $\lambda$21\,cm signal. The latter requires an interferometer with thousands of square meters of collecting area to have adequate sensitivity, whereas a single dipole could have sufficient sensitivity to detect the sky-averaged spectrum. In either case, the principle challenge arises from foreground emission that is at least four orders of magnitude brighter than both the anticipated angular fluctuations \citep[i.e.,][]{bernardi09,bernardi10,ghosh12,pober13,paciga13} and the sky-averaged signal \citep[i.e., ][]{deoliveiracosta08,rogers08}.  Previous works have relied on the anticipated spectral smoothness of the foregrounds in frequency to separate them from the less-smooth 21cm signal:  The 21cm signal should vary over kHz scales in pencil--beam observations, and the sky--averaged 21cm signal is predicted to show variations over scales of $\sim$10~MHz.  
However, it is thought that the foregrounds follow an approximate power-law, with deviations on scales much larger than 10~MHz (a contention investigated here).

Recently, measurements of the sky-averaged signal from the EoR and earlier epochs have received renewed attention owing to limits placed on reionization models by the Experiment to Detect the Global EoR Signature \citep[EDGES,][]{bowman10}. This has inspired several theoretical investigations of the constraining potential and optimal survey/analysis strategies for measurements \citep{pritchard10,harker12,morandi12,liu12,switzer14}, as well as  new detection efforts. 
The Shaped Antenna measurement of the background RAdio Spectrum \citep[SARAS;][]{patra12} project targets the EoR whereas   
the Large-aperture Experiment to detect the Dark Age \citep[LEDA;][]{greenhill12, taylor12} the LOFAR Cosmic Dawn Search \citep[LOCOS;][]{vedantham13} and SCI--HI \citep{voytek14} target the transition era between the dark age and EoR ($z\sim 20$).  Finally,  the Dark Age Radio Explorer \citep[DARE;][]{burns11} space mission concept is intended to enable study of the sky-averaged $\lambda21$\,cm signal from the EoR up to $z \sim 30$.

This paper focuses on the detection of the sky-averaged $\lambda$21\,cm signal from the neutral IGM at the close of the dark age and beginning of the EoR.  This epoch is forecast to appear as an absorption trough much greater in magnitude than the emission feature associated with the EoR in current theoretical models \citep[]{furlanetto06a,pritchard10}.  The trough morphology is determined by (1) the onset, the strength, and the evolution of coupling to the Ly$\alpha$ background from the first generations of stars and (2) the heating by X-rays from byproducts of stellar evolution such as supernovae and black holes \citep[i.e.,][]{furlanetto06a,fialkov14,mirocha14}.  Other processes such as dark matter annihilation \citep[]{valdes07,valdes08,valdes10,valdes13} or more exotic mechanisms (e.g., \citealt{mack08}) may also affect the amount of $\lambda21$cm absorption. 

We study the effect of antenna response on the apparent frequency spectrum of foreground emission. \cite{vedantham13} were the first to include the effect of a broadband antenna gain pattern and showed that, in the case of LOFAR dipoles, angular and spectral response has a considerable effect.  We expand on the results of this initial investigation in several ways:
(1) we develop a more physical understanding of how the spectrum and angular distribution of the foregrounds impact sky-averaged measurements in a realistic instrument [where the foreground structure couples to the angular and spectral response of the antenna].  (2) After finding that the instrumental response is the limiting factor for sky-averaged spectra, we explore the benefits of calibrating the antenna gain pattern through interferometric measurements.   (3) Previous studies did not quantify how well the foregrounds must be modeled to be able to detect the HI signal.  We forecast the model complexity needed to yield an unbiased detection.  

In outline, Section~\ref{sec:instrument} first describes our simulations of the instrument and foreground sky and then investigates how well different foregrounds models can be subtracted as a smooth polynomial both for ideal instruments and for when a realistic antenna gain pattern is assumed. Section~\ref{leda_approach} shows how calibration of the dipole gain pattern used to detect the global signal through an interferometric array helps the foreground modeling and subtraction. First-order forecasts of effectiveness for LEDA at constraining the cosmological signal are presented in Section~\ref{sec::fisher}, with conclusions thereafter.\\

\section{Antenna Response and Sky Brightness}
\label{sec:instrument}

At time $t$ and in pointing direction $\hat{n}$, an individual antenna provides a measurement of the beam--averaged sky brightness temperature,\\
\begin{eqnarray}
T(t,\nu, \hat{n}) &=& T_{\rm N}(t,\nu)+ \left({\int_\Omega A_{\hat{n}}(\nu,\hat{n}') \; d \hat{n}'} \right)^{-1} \nonumber \\
&\times& { \left[\int_\Omega T_{\rm sky}(t,\nu,\hat{n}') A_{\hat{n}}(\nu,\hat{n}') \; d \hat{n}' \right] g(t,\nu)}, 
\label{dipole_eq}
\end{eqnarray}

\noindent
where $\nu$ is frequency, $T_{\rm sky}$ is the direction dependent sky brightness temperature, $A_{\hat{n}}(\nu,\hat{n}')$ is the antenna gain pattern in the direction $\hat{n}'$, and $g$ represents the overall receiver gain. The noise due to the receiving system, $T_{\rm N}$, is dominated by the sky noise at frequencies corresponding to the $\lambda21$cm signal from $z\sim20$ and is, therefore, affected by the antenna field of view.   Many of the ensuing calculations will investigate how well the foregrounds can be modeled for a noiseless, ideal receiving system.  However, the impact of noise can easily be understood as it will enter as a constant rms term that does not significantly decrease with increasing polynomial fitting order (see Section~\ref{foreground_simulation}).

Ignoring environmental effects (e.g., variable soil moisture), the antenna gain pattern is time-independent, and the observed sky spectrum after time integration $\Delta t$ is
\begin{eqnarray}
T(\nu, \hat{n})  = \int_t^{t+\Delta t} T(t,\nu, \hat{n}) \, d t.
\label{integrated_spectrum}
\end{eqnarray}

\noindent
We can decompose $T$ into components from foregrounds and $\lambda$21\,cm radiation
\begin{eqnarray}
T(t,\nu,\hat{n}) = T_{f}(t,\nu,\hat{n})  + T_{\rm HI}(\nu),
\label{sky_decomposition}
\end{eqnarray}
separating terms with different dependancies on time, frequency, and position.

The present work seeks to test whether the $\lambda$21\,cm signal can be distinguished from the foreground signal, and up to what order in the foreground brightness model, where  $\log T_f$ is expressed as a polynomial in $\log\nu$ such that
\begin{eqnarray}
\log \hat {T}_{f}(\nu) = \sum_{n = 0}^m c_n \/ (\log{\nu})^n.
\label{pol_model}
\end{eqnarray}
The hat symbol denotes a modeled quantity. This polynomial form is a common foreground parametrization \citep[]{pritchard10,bowman10,harker12}, motivated by the power law--like distribution of the non-thermal cosmic ray electrons that dominate the emission at frequencies ${\cal O}(100)$\,MHz, via the synchrotron process, and the apparently power-law spectra of most sky structures that contribute to foreground emission.  \citet{liu12} and \citet{vedantham13} discuss using a principle component approach to describe foregrounds. In practice, this  is complicated because the foregrounds and $\lambda$21\,cm signals are superposed \citep{vedantham13}, and when a single antenna is used, separation reliant on the angular variation of the foreground signal as advocated in \citet{liu12} is not possible.   We do not attempt to address whether there is a better basis to subtract foregrounds here, but we note that two other simple choices of model $\hat {T}_{f}(\nu)$, sinusoids and polynomials that are normal in $\nu$, do not yield residuals dominated by the cosmological signal after subtraction for {\it any} model order.

In the $\log{\nu}$ space, subtraction is not a linear operation, so (formally) different components of $T(\nu)$ (i.e., foreground and the $\lambda$21\,cm signal) cannot be separated readily. 
However, because the $\lambda$21\,cm signal is much smaller than the foreground contribution, one can Taylor expand $\log T$:
\begin{equation}
\log T \approx \log T_f + T_{\rm HI} / T_f.
\label{eqn_poly}
\end{equation}
Therefore, one can separately consider how subtraction fares on the two components of ${T}(\nu)$. 

In what follows, we use the RMS of the residual spectrum, as a function of polynomial order, as a metric for the effectiveness of foreground subtraction, i.e., 
\begin{equation}
RMS_{\rm res} = \sqrt{\langle (T(\nu) - \hat T(\nu))^2 \rangle},
\end{equation}
where $\langle ... \rangle$ indicates an average over frequency.  
In selected cases we also plot the actual residuals as a function of order to show their behaviour across the frequency band.

\begin{figure}
\begin{center}
\epsfig{file=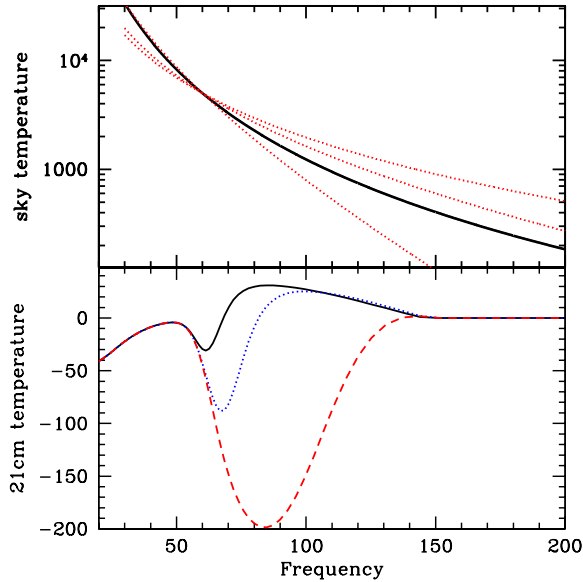, width=8.3cm}
\end{center}
\caption{ ({\it Top)}-- Synchrotron signal from a mono-energetic electron populations with energies $E = 0.5$, $1$, and $2~$GeV (red dotted curves), which peak in total emitted power for a single electron at $\nu_c =10$, 50, and 200\,GHz, respectively, and for a physically-motivated diffusive model of Galactic cosmic ray electrons (solid black curve). See Section~\ref{sec:simple_models} for additional description.  ({\it Bottom})--  Three models for the sky-averaged $\lambda$21\,cm signal computed using the parametrization described in \citet{furlanetto06a} and \citet{mcquinn12}.  The differences arise primarily due to the amount of X-ray heating of neutral gas by supernovae and X-ray binaries.  All models assume the same reionization history.
\label{fig:global} }
\end{figure}

\subsection{Antenna Gain Pattern Models}
\label{beam_models}

\begin{figure}
\begin{center}
{\epsfig{file=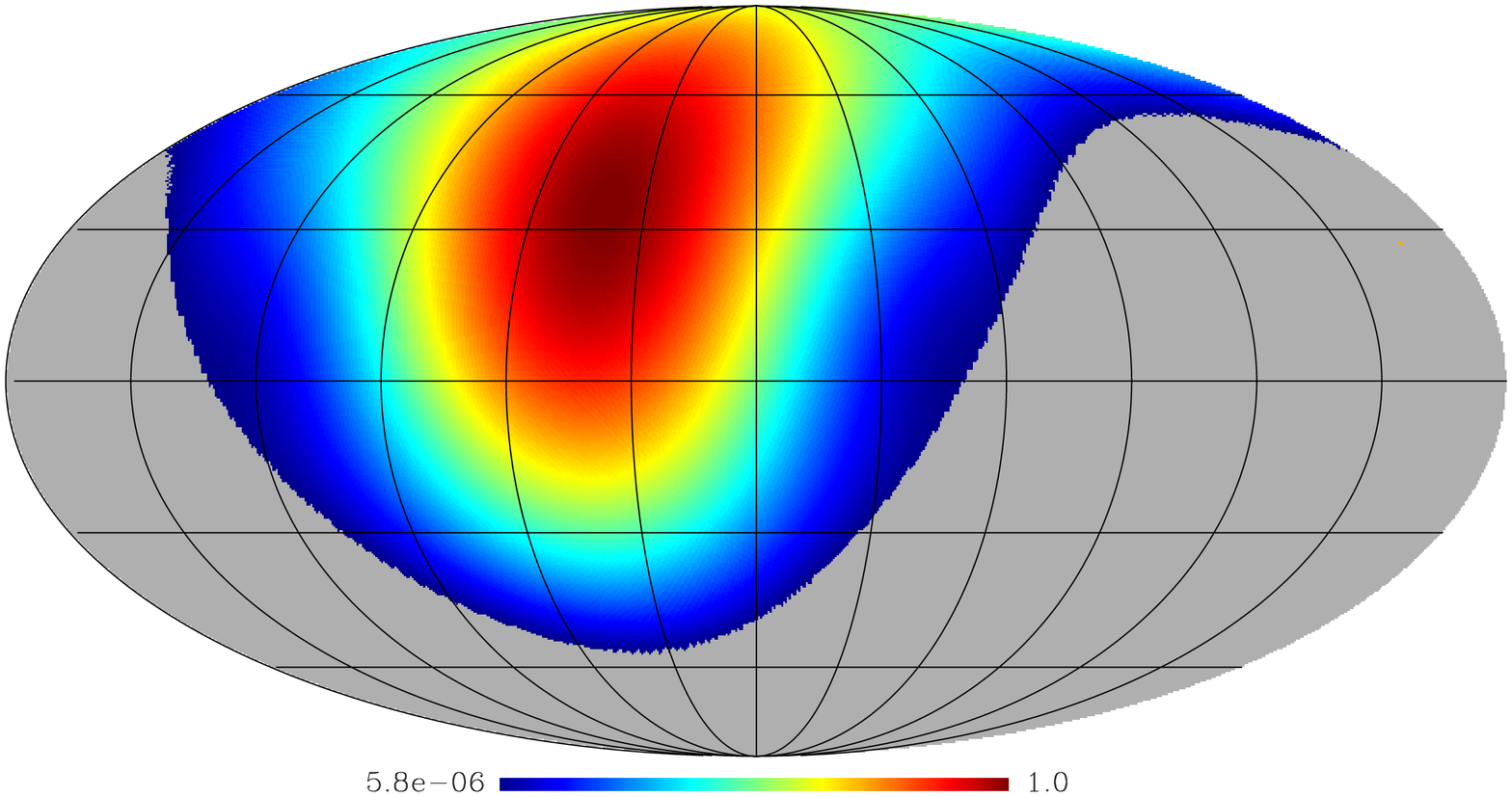,angle = 90, width = 8.5cm}}
{\epsfig{file=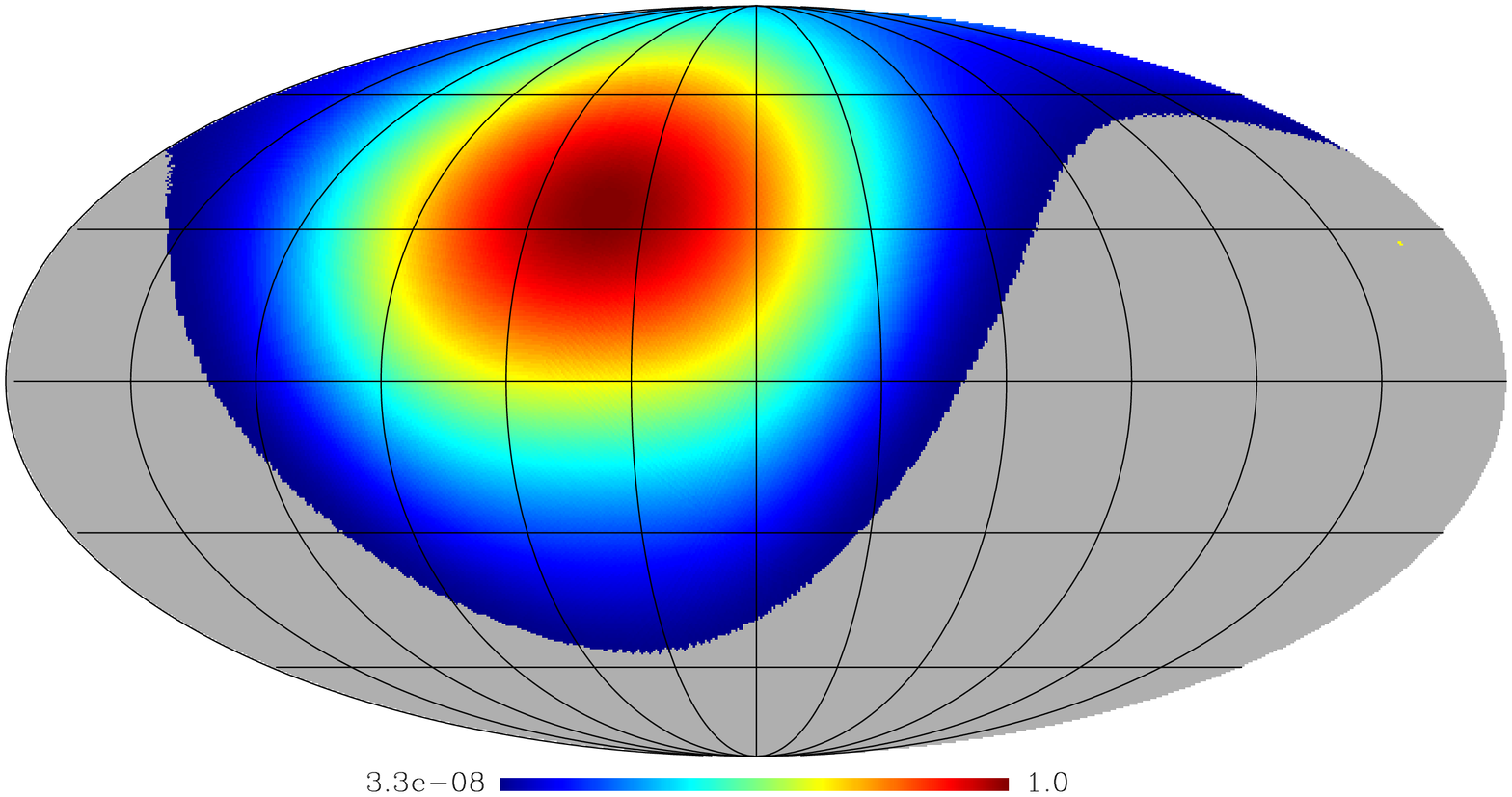,angle = 90, width = 8.5cm}}
\end{center}
\caption{Antenna gain patterns  at 40\,MHz and a single sidereal time:  simple short dipole (top) and LWA inverted-V dipole (bottom).  Gray represents the sky below the horizon (equatorial coordinates). The scale is logarithmic (color bars)\label{fig:dipoles}}
\end{figure}

In order to simulate the effect of nonuniform instrument response on the sky spectrum, 
we test two models for the antenna gain pattern, $A(\nu,\hat{n})$. The first is that of a simple short dipole over a ground plane \citep{kraus50}:
\begin{eqnarray}
A_{\rm d}(\nu,\theta,\phi) = 2 \sin \left(2 \pi h \cos{\theta} \right) \sqrt{1 - (\sin{\theta} \sin{\phi})^2}, 
\end{eqnarray}
where $\theta$ and $\phi$ are the zenith and azimuth angles, respectively,  and $h$ is the dipole height above the ground plane, expressed in wavelength.  Variation with frequency, mediated entirely via $h$, is slow.  The second model is that for the broadband inverted-V dipole used in LEDA and provided by the Long Wavelength Array \citep[LWA,][]{taylor12, ellingson13}.  This may be parametrized as follows:
\begin{eqnarray}
A_{\rm LWA}(\nu, \theta, \phi)\hspace{-0.03in} =\hspace{-0.03in} \sqrt{[p_E (\nu,\theta) \, \cos \phi]^2 \hspace{-0.03in} + \hspace{-0.03in} [p_H (\nu,\theta) \, \sin \phi]^2},
\end{eqnarray}

\noindent
where $E$ and $H$ represent the two orthogonal polarization axes for a single dipole and 

\begin{eqnarray}
p_i (\nu,\theta) & = & \left[ 1 - \left( \frac{\theta}{\pi/2} \right)^{\alpha_i(\nu)} \right] \, (\cos{\theta})^{\beta_i(\nu)} \, + \\ \nonumber
	    & & \, \gamma_i(\nu) \, \left ( \frac{\theta}{\pi/2} \right ) \, (\cos{\theta})^{\delta_i(\nu)},
\label{lwa_dipole_model}
\end{eqnarray}
with $i=E,H$  \citep{ellingson10}. Estimates of the  exponents were obtained using a semi-analytic propagation model.  Following  \citet[]{dowell11}, we approximate the vector ${\bf a}_i(\nu) \equiv [\alpha_i(\nu), \beta_i(\nu), \gamma_i(\nu), \delta_i(\nu)]$ for each plane using  a polynomial in frequency,
\begin{eqnarray}
a_{i,j}(\nu) = \sum_{n = 0}^m a_{i,j}^{(n)} \/ \left ( \frac{\nu}{\nu_0} \right )^n,
\label{lwa_dipole_frequency_model}
\end{eqnarray}
where $j$ enumerates the vector element ($\alpha$,...,$\delta$), finding suitable fits over 40-88\,MHz (the approximate LEDA passband) in both planes for $m=3$ (Figure~\ref{fig:dipoles}). As in  \citet{ellingson10}, $[\gamma_H(\nu),\delta_H(\nu)]$ is fixed at zero. 

\subsection{Sky Models}
\label{sec:simple_models}

We consider models where the spatial structure of the foregrounds is crudely captured, but the properties of the foregrounds (and hence their frequency dependence) are physically modeled. In the process, we assess the impact that the most pessimistic assumptions about the foreground frequency spectrum  have on the detection of the $\lambda$21\,cm spectrum.    

\subsubsection{Diffuse Emission}
\label{sec:matt_models}
Galactic synchrotron emission is believed to be the dominant foreground  below 100 MHz (e.g., \citealt{dimatteo02, santos05, mcquinn06}), so we begin by modeling this component. A more general model that includes the small contribution from extragalactic sources follows. For measurement of {\it fluctuations} in the $\lambda$21\,cm signal, foreground removal for an idealized experiment is almost lossless even for a mono-energetic population of synchrotron electrons \citep{petrovic11}, ignoring the effect of mixing angular and spectral modes introduced by the instrument \citep[][]{datta10,vedantham12,trott12,morales12,parsons12,thyagarajan13}.  This result suggests that even the most pathological synchrotron foreground imaginable does not necessarily impede detection of the $\lambda$21\,cm fluctuations signal.  We investigate here whether the same holds for the sky-averaged signal. 

The spectrum of power emitted by a single electron via the synchrotron process is \citep[e.g.,][]{rybicki79}
\begin{equation}
P(\nu) \propto \frac{\nu}{\nu_c} \int_{\nu/\nu_c}^\infty K_{\frac{5}{3}}(\xi) d\xi,
\end{equation}
where $K$ is a modified Bessel function of the second kind, $\nu_c \propto E^2 B_\perp$, $E$ is the energy of the electron, and $B_\perp$ is the perpendicular component of the magnetic field. The mapping between the peak of $P(\nu)$ and electron energy is
\begin{equation}
E \approx \left( \frac{\nu}{\rm 50\,MHz} \right)^{1/2} \left( \frac{B_\perp}{10 ~\mu {\rm G}}\right)^{-1/2}~~{\rm GeV},
\end{equation}
where $10\,\mu {\rm G}$ is characteristic of the Galactic magnetic field.  The three red dotted curves in the top panel of Figure\,\ref{fig:global} show the brightness temperature of the synchrotron spectrum from a mono-energetic electron population with energy of $0.5$, $1$ or $2$\,GeV and $B_\perp = 10\,\mu G$, normalized to $5000$\,K at $60$\,MHz. The $1$\,GeV case has peak frequency at $\nu_p = 50$\,GHz, and the other cases fall a factor of $4$ above and below this $\nu_p$. Even the synchrotron emission from a mono-energetic electron population presents a smoother spectrum (and with fewer inflection points) than models of the $\lambda$21\,cm signal (bottom panel of Figure\,\ref{fig:global}).

In reality, the sky-averaged Galactic synchrotron spectrum arises from the summation of $P(\nu)$ over a broad distribution of electron energies.  We explore a minimal model for the Galactic electron population to assess how readily the sky-averaged $\lambda 21$\,cm signal can be extracted in a case that approximately represents cosmic ray electrons.  In particular, we solve the diffusion equation for a disk with a cosmic ray source profile of $\exp(-0.5 [r/8{\rm~ kpc}]^2) \delta^D(z)$ where $r$ is Galactocentric radius and $z$ is the height above the mid plane (to emulate the distribution of supernovae), a 3\,kpc  diffusion `halo' of height  above the disk with vacuum boundary conditions, a diffusion coefficient $D_0 = 3\times10^{28}~{\rm cm^2 s^{-1}}$, and an electron energy loss rate of 
\begin{eqnarray}
\frac{d\log \gamma}{dt} &=& [6\times10^{-13} \, n_{HI}  + 10^{-15} \,n_e  \, \gamma + \nonumber \\
                        &+& 1.2 \times 10^{-19} \,\gamma^2 ] {\rm ~ s^{-1}},
\label{eqn:dEdt}
\end{eqnarray}
after \citet{atoyan95}.  The three cooling terms in the bracket from right to left stem from losses owing to ionization, Coulomb collisions, and synchrotron emission plus inverse Compton. Here $n_{HI}$ and $n_e$ are the number densities of hydrogen atoms and electrons, respectively, in units of cm$^{-3}$ (set to unity in our calculations), $\gamma = E/m_e c^2$, and the last factor on the RHS assumes $B_\perp = 10~{\rm \mu G}$ and the radiation energy density is dominated by the cosmic microwave background.  This model is similar to what is used in diffusion models such as GALPROP,\footnote{\url{http://galprop.stanford.edu/}} which is able to match many of the observed properties of Galactic cosmic rays.  We integrate the resulting electron energy distribution from this calculation over the synchrotron kernel, $P(\nu)$, to generate the model sky spectrum.  Collisional and ionization cooling start to become relevant at energies $<1$\,GeV.  Interestingly, below 1~GeV -- electrons that are important for the low frequency radio emission -- collisional/ionization cooling starts to become relevant as opposed to inverse Compton and synchrotron cooling at higher energies.  This produces a break in the cosmic-ray electron energy spectrum, characterized by a transition from $dN/dE \sim E^{{-\beta-1} }$ to $E^{{-\beta+1} }$ once ionization losses become important, where $\beta \approx 2.2$ is the injection index of accelerated electrons.  This break is in agreement with more sophisticated Galactic diffusion models (e.g., \citealt{strong07}).\footnote{There are two other effects that we do not model but that could also impart additional structure in the radio.   There is some evidence for an intrinsic break at  $\sim10$~GeV in the injected spectrum from supernovae \citep{strong00}.  In addition, re-acceleration of cosmic rays off waves in the interstellar medium may start to become important at $E \lesssim 1$\,GeV \citealt{strong07}).}  

The diffusion model results in a synchrotron spectral index of $\approx -2.7$ over 100-200\,MHz, changing by only a percent between 30 and 120\,MHz (Figure\,\ref{fig:global}, top panel, black curve), and in rough agreement with the measurement of \citet{rogers08}.  As this model incorporates a minimal amount of physics, it is likely to be smoother than the true Galactic synchrotron spectrum (although, we have assumed only one value for $B_\perp$ and for the densities in equation~(\ref{eqn:dEdt}), and dispersion in these values will also act to smooth the spectrum further).\\

\begin{figure}
\begin{center}
\epsfig{file=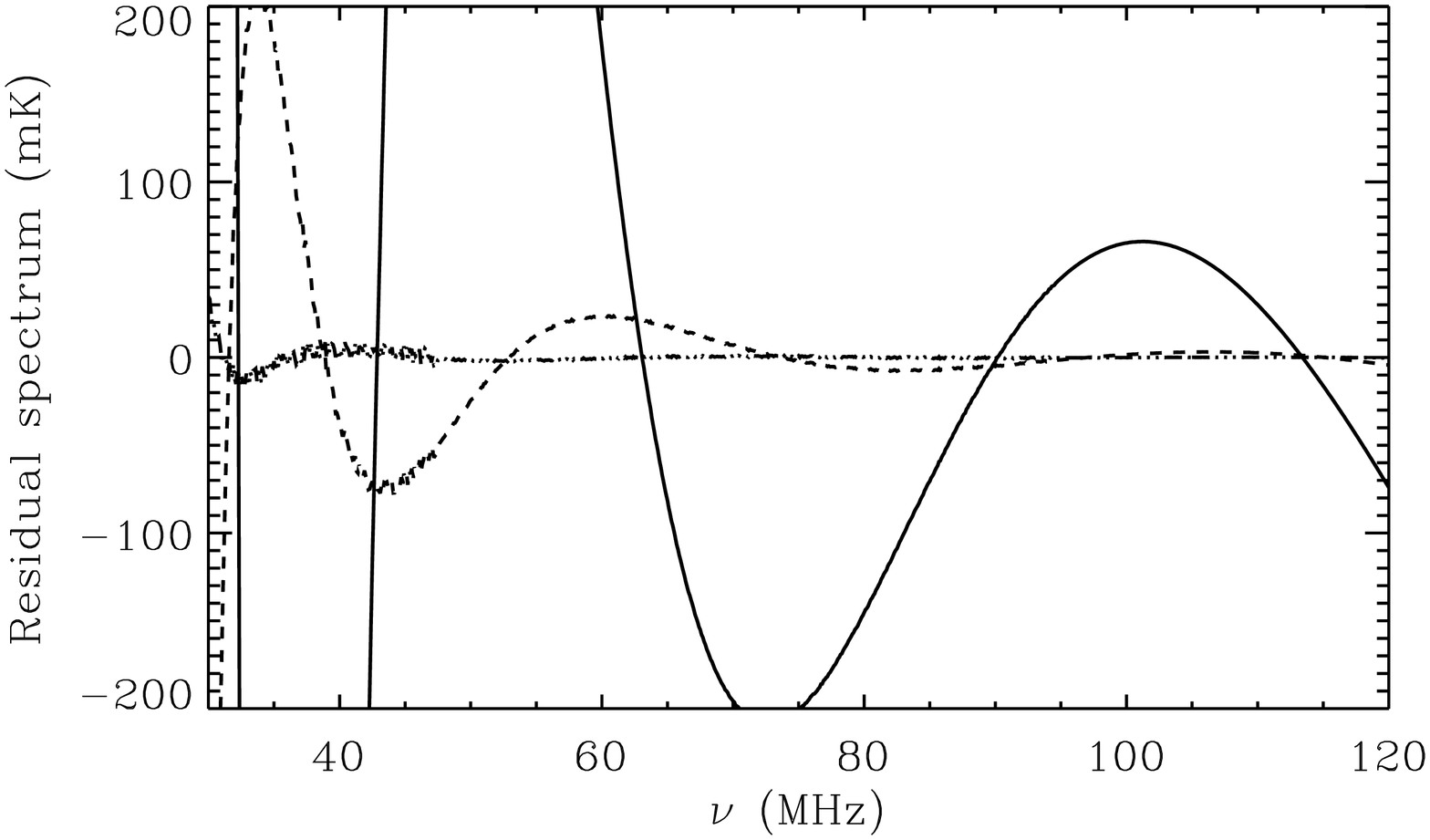, angle = 90, width=8.5cm}
\epsfig{file=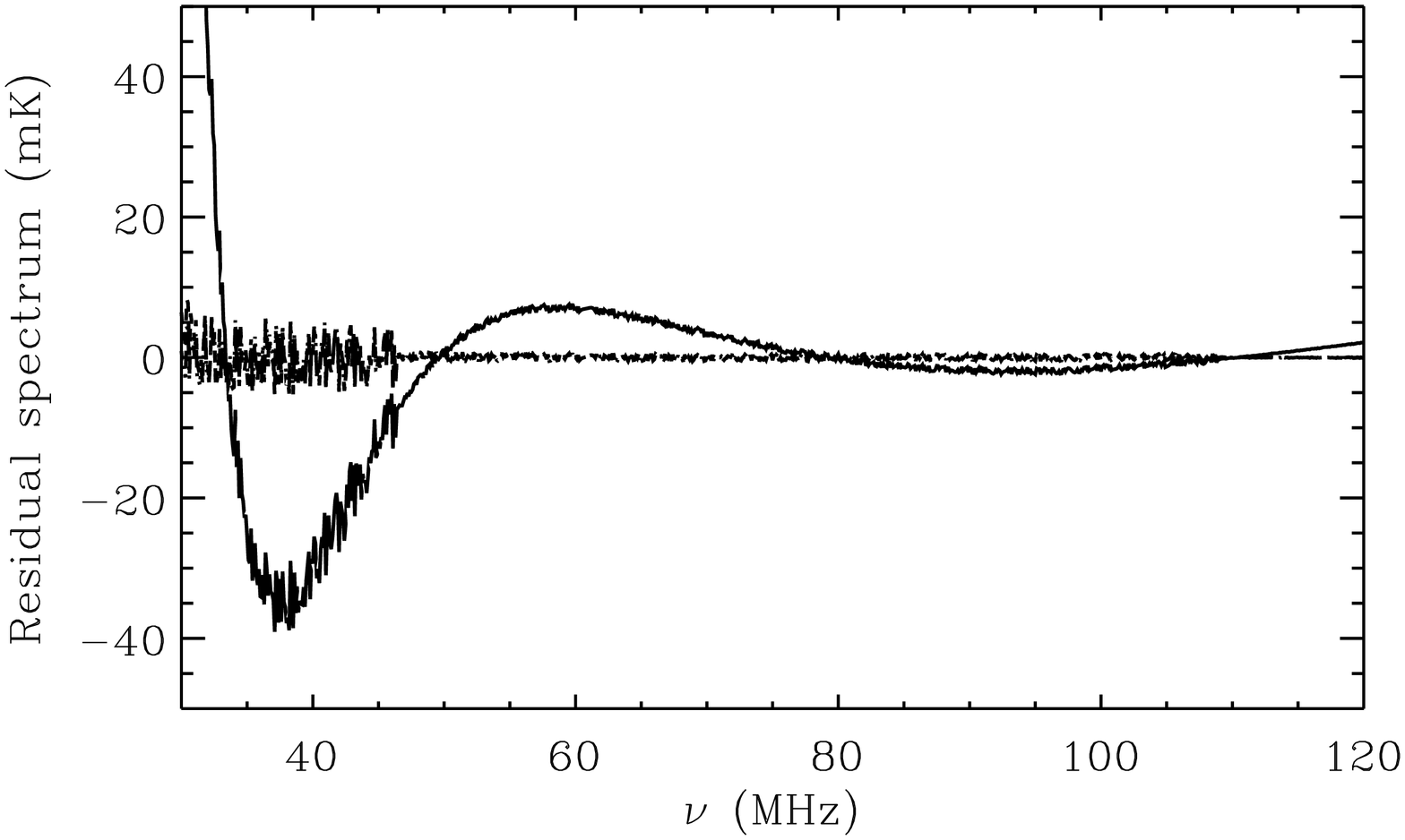, angle = 90, width=8.5cm}
\end{center}
\caption{ ({\it Top)} -- Residuals after subtracting a $4^{\rm th}$ (solid line), a $5^{\rm th}$ (dashed line) and a $6^{\rm th}$ (dot-dashed line) order polynomial in $\log\nu$  (Eq. \,\ref{pol_model}) for the case of a mono-energetic, $1$\,GeV electron population.  The residuals decrease with increasing order.  ({\it Bottom)} --  Same as the top panel but for a $3^{\rm rd}$ and a $4^{\rm th}$ order polynomial and for a synchrotron foreground model that uses the minimal diffusion model for the cosmic ray electron population. The models in the top and bottom panels can be thought of as the most pessimistic and optimistic cases, respectively. The implication is that higher-order polynomials than considered in previous studies are required when subtracting the foreground synchrotron emission. The increased residual noise below $\sim$45~MHz is due to numerical artefacts in the model interpolation.\label{fig:projections}}
\end{figure}

We find that a higher-order polynomial in $\log\nu$  is required to represent the synchrotron spectrum than has been considered in previous studies such as \citet{pritchard10} and \citet{harker12}.  Figure~\ref{fig:projections} shows the residuals after subtracting different order polynomials in $\log\nu$ over 30-120\,MHz (comparable to the LEDA passband) for our different models for the synchrotron spectrum. The top panel shows the contrived case of a $\delta$-function electron distribution with $E=1$\,GeV, which represents the most pessimistic case possible as not only it is mono-energetic but the break in the synchrotron spectrum occurs in the middle of the band for this energy. This pathological case requires a $6^{\rm th}$ order polynomial for the residuals to be less than the anticipated amplitude of the $\lambda21$\,cm signal, $\sim 100$\,mK (Figure\,\ref{fig:global}).  The bottom panel shows the sky spectrum in the more realistic diffusion model for the distribution of Galactic cosmic ray electrons.  In this case, at least a  $4^{\rm th}$ order polynomial is required to reduce residuals below $100$\,mK. We have also investigated fitting over a broader band, 30-200\,MHz and find that an additional order is required to yield similarly small residuals. We note that at higher frequencies (i.e., the EoR), it is possible that a somewhat lower order polynomial may be used owing to reduced sky brightness, but on the other hand, higher orders may be demanded by the smaller magnitude of the signal.

\subsubsection{Point Source Emission} 
Extragalactic point sources constitute $\sim 10\%$ of sky brightness temperature \citep[i.e.,][]{dimatteo02,jackson05,jelic08}.  Let us assume that the intensity and spectral index of the point sources is uncorrelated and that the distribution of spectral indices is a Gaussian with width $\sigma_\alpha$ at reference frequency $\nu_0$.  Then, the result of convolving the background with a Gaussian distribution of spectral indexes with width $\sigma_\alpha$ is
\begin{equation}
T = T_0 \left( \frac{\nu}{\nu_0} \right)^{\bar{\alpha} + \bar{\alpha} \, \sigma_\alpha^2\log(\frac{\nu}{\nu_0})/2},
\end{equation}
where $T_0$ is the average brightness temperature at $\nu_0$ and we assumed $\sigma_\alpha \approx 0.3$ \citep[i.e,][]{mauch03}.
Thus, at least for this simple case, a  $2^{\rm nd}$order polynomial in $\log \nu$ is sufficient to remove the foregrounds, but any deviations from a Gaussian spectral index distribution (or power-law functional forms) will impart higher order terms.  For example, a skewness  (defined as the  $3^{\rm rd}$ moment of the distribution in $\alpha/\sigma_\alpha$) adds  $3^{\rm rd}$ and  $5^{\rm th}$ order terms (with the latter suppressed by an additional $\sigma_\alpha^4$).  Kurtosis (defined as one higher moment) adds $4^{\rm th}$ and  $6^{\rm th}$ order terms.  For a $\sim 10$\% contribution to sky temperature, skewness at $1:10^3$ would require fitting these extra terms to reduce residuals by  ${\cal O}(10^4)$ over a band with $\Delta \nu \sim \nu_0$.  Since a skewness at this level is almost inevitable, removing point sources with different spectral indices also requires fitting at a similar order to what the earlier diffusion model suggests is necessary in order to remove Galactic synchrotron emission.

\subsection{Angular Structure}
\label{foreground_simulation}

\begin{figure}
\begin{center}
{\epsfig{file=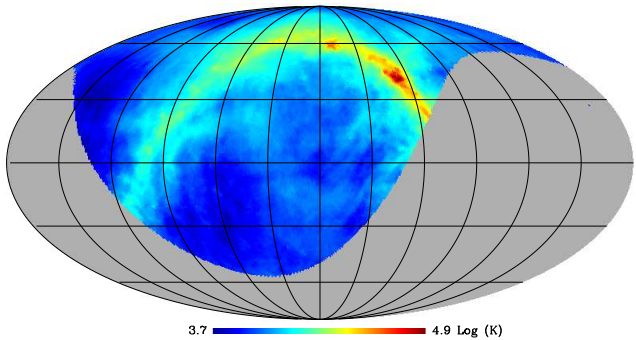, angle = 90, width=8.5cm}}
\end{center}
\caption{Foreground simulated map at 40~MHz. The gray area is below the horizon at LST = 7~hours for $+34^\circ$ latitude.\label{fig:galactic_model}}
\end{figure}

Coupling between angular and frequency structure in antenna gain patterns and foreground emission may be anticipated to exacerbate residuals unless models include high order polynomial terms. We investigated this problem here by adopting two reference maps of foreground emission: the all-sky 408 MHz map \citep[]{haslam82} and a 150~MHz all-sky map from the \cite{deoliveiracosta08} model. We derived a direction-dependent spectral index from the ratio of 150 and 408~MHz temperatures and extrapolated the 150~MHz map below 100 MHz (e.g., Figure~\ref{fig:galactic_model}). We will refer to this foreground model as our ``foreground simulations". Albeit smoother in frequency, our model accounts for angular variations in the sky spectrum that were not accounted for in the all-sky principle component analysis of \cite{deoliveiracosta08}.

We simulate observations using an antenna at $+34^\circ$ latitude (which corresponds to the LEDA prototyping site at the LWA in New Mexico. The results are substantially similar to those for the $+37^\circ$ latitude LWA site in the California Owens Valley at which the full LEDA system has been deployed).
We model how the sky brightness distribution multiplied by the gain pattern varies as a function of local Sidereal Time for $7^{\rm h}$ - $10^{\rm h}$.  The Galactic Center is below the horizon during this interval.  We created four \emph{noiseless} sky realizations evaluated on the hour (Figure~\ref{fig:galactic_model}), constructed  products with gain pattern models over 1\,MHz bins, integrated in angle and averaged in time.  These mock observations account for the coupling between angular and frequency structure in the foreground sky and antenna gain patterns.  We note that actual observations would use higher time cadence and frequency resolution so as to enable excision of  radio frequency interference and calibration of gain fluctuations in the signal path prior to averaging.

\begin{figure}
\begin{center}
{\epsfig{file=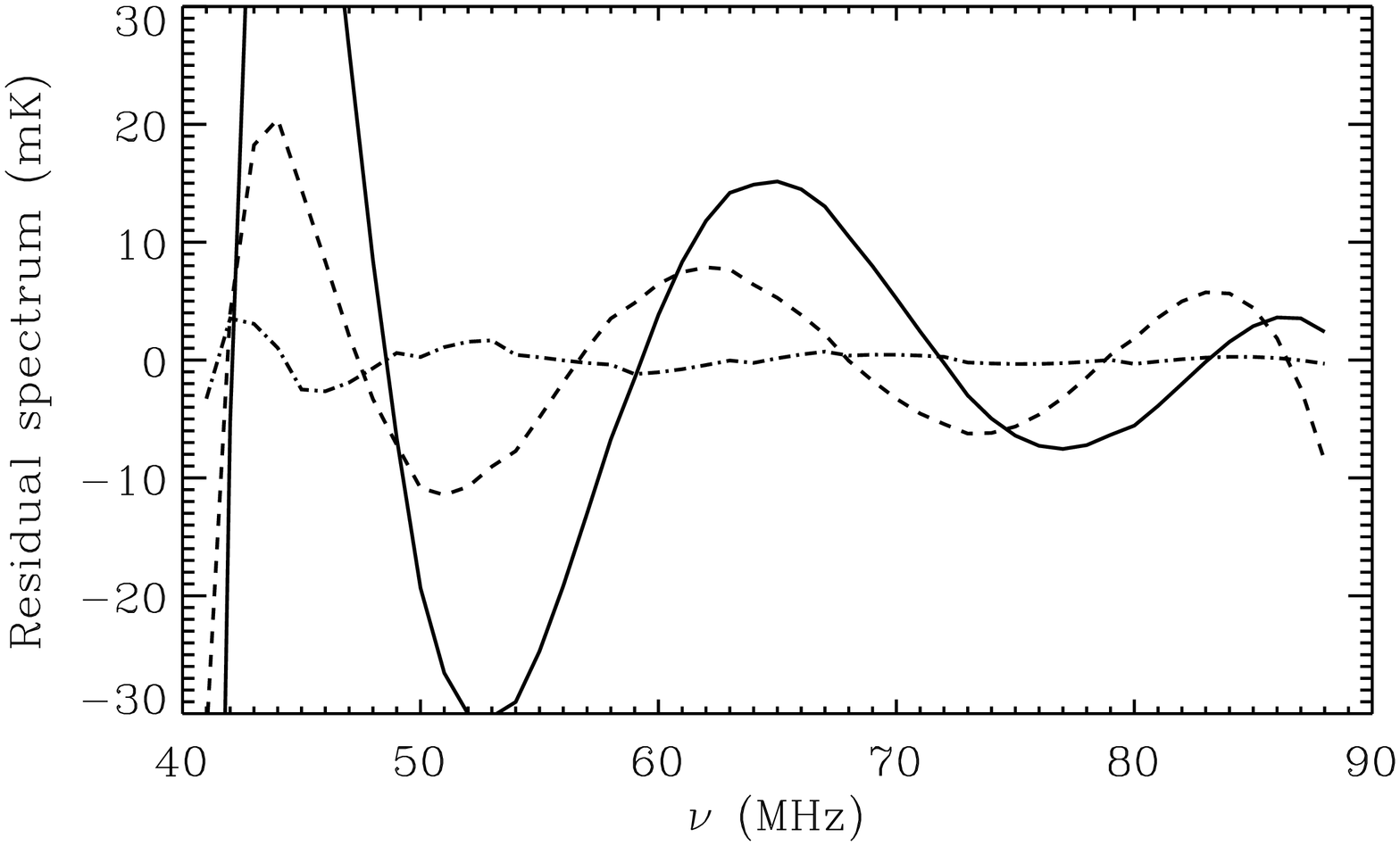, angle = 90, width=8.5cm}}
{\epsfig{file=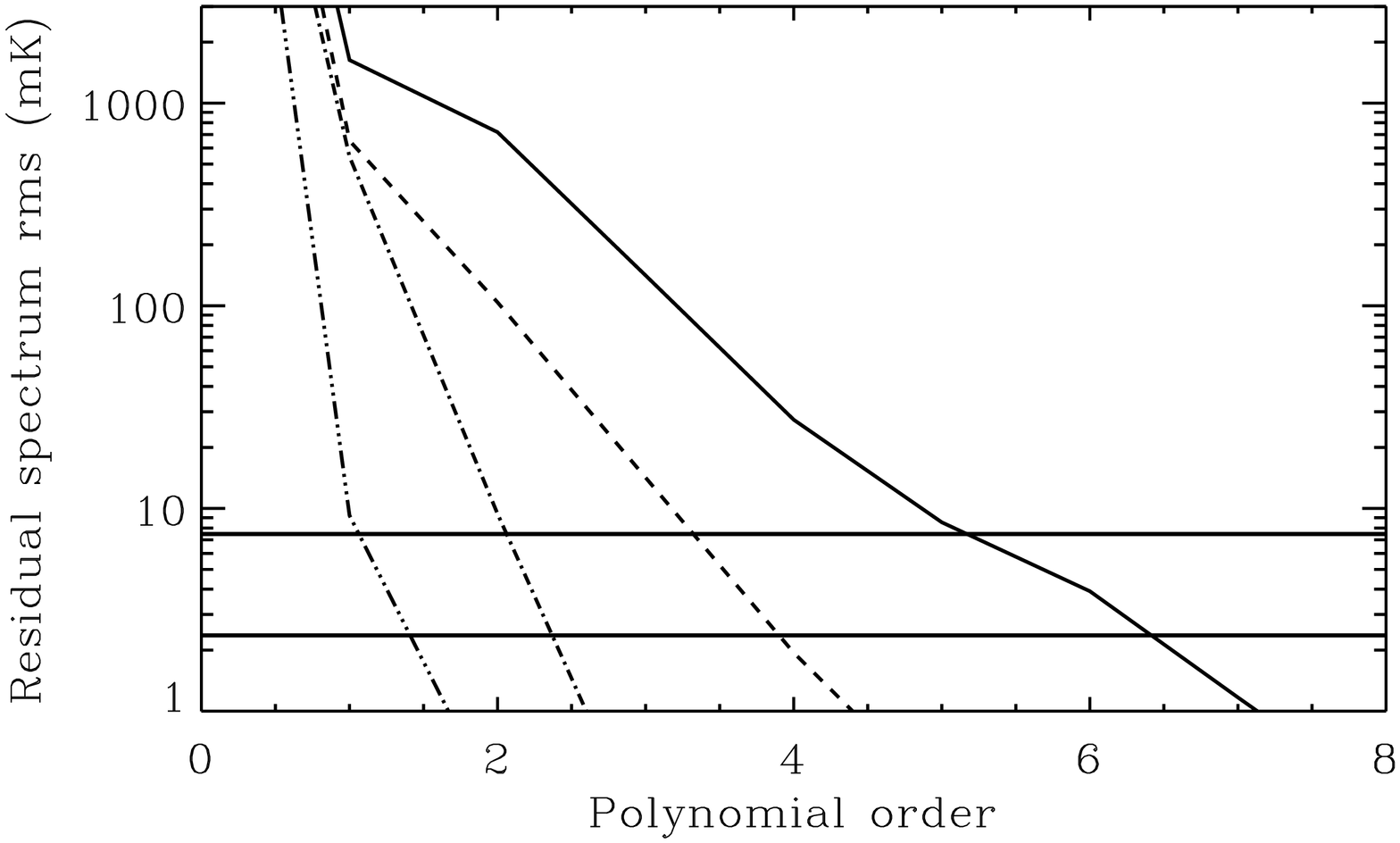, angle = 90, width=8.5cm}}
\end{center}
\caption{({\it Top)} --  Spectrum of residuals for the sky model described in Section~\ref{foreground_simulation} and the gain pattern of an LWA dipole.   Polynomials in $\log\nu$ for a 4$^{\rm th}$, 5$^{\rm th}$ and 7$^{\rm th}$ orders (solid, dashed and dot-dashed line respectively) are subtracted from the foreground model, averaged in angle and time.  ({\it Bottom)} -- RMS over 40-88\,MHz after polynomial subtraction for four cases: the foreground simulation alone (i.e., no multiplication by a dipole gain pattern, triple dot--dashed line); the minimal cosmic ray electron diffusion model (Section~\ref{sec:simple_models}), which interestingly yields more frequency structure though it carries no intrinsic angular structure (dot-dashed line);  foreground simulation multiplied by the short dipole gain pattern (dashed line); and the same multiplied by the LWA dipole pattern (solid line). The horizontal lines are the RMS noise for four single dipoles observing for 10 and 100\,hr.\label{fig:residual_fit}}
\end{figure}

Next we subtracted fitted polynomials as described by Equations~\ref{pol_model} and \ref{eqn_poly}. Residuals as functions of frequency and polynomial order are shown in Figure~\ref{fig:residual_fit}. For order $n=2$ and the foreground simulations without multiplication of a dipole response, the RMS residual is $\sim 1$~mK. This reflects the smoothness of the intrinsic sky model to such a degree that the sky averaged $\lambda$21\,cm signal would be readily detectable.

Incorporation of a short dipole gain pattern coupled to the sky raises RMS residuals above 1\,mK for $n<5$.  For an LWA dipole, the same is true for $n<7$.  
We conclude that the frequency structure in observed residual spectra is more sensitive to the detailed angle and frequency dependence of antenna gain pattern than that in the extant sky model. 
This persists for foreground simulations that include the minimal cosmic ray electron diffusion model (Section~\ref{sec:matt_models}), for which variation with frequency is more complex  but the effect is homogeneous across the sky - residuals are $>1$\,mK for $n < 3$.

\begin{figure}
\begin{center}
{\epsfig{file=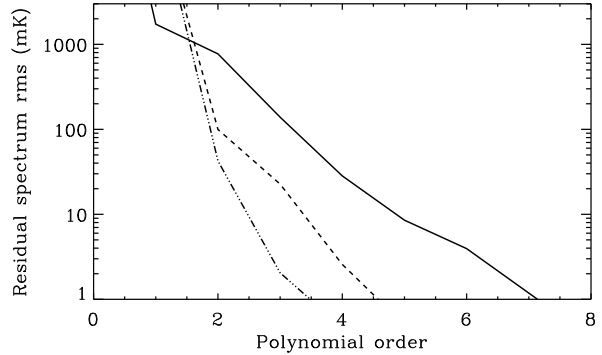, angle = 90, width=8.5cm}}
\end{center}
\caption{Same as the bottom panel of Figure~\ref{fig:residual_fit} except that normally distributed variation in the spectral index on $7^\circ$ scales with a 0.5 RMS has been added to the foreground simulation. It is notable that with accounting for the LWA gain pattern,  there is little difference between residuals with and without dispersion in spectral index (Figure~\ref{fig:residual_fit}). 
\label{fig:residual_fit3}} 
\end{figure}

In order to test for robustness, we considered a foreground model in which Gaussian fluctuations are added to the spectral indices derived in Section~\ref{foreground_simulation} pixel by pixel and convolved with a  $7^\circ$ FWHM Gaussian kernel in order to preserve angular coherence. We derived foreground models and residual spectra for a 0.5 RMS in spectral index, on a $7^\circ$ scale (Figure~\ref{fig:residual_fit3}). This can be considered a worst case since spectral index measurements with similar resolution at $\sim$1\,GHz have peak-to-peak variations of 0.1 \citep{reich88}.
We observe that the primary impact on the RMS of the residual spectrum is to increase the required polynomial order for an antenna with uniform response.  Otherwise the impact of the antenna gain pattern dominates and so the  foreground model that accounts for varying spectral indices will not be used in later sections.

\section{Aid from an interferometric array}
\label{leda_approach}

Extraction of the $\lambda$21\,cm signal from spectra requires establishing a high-order polynomial representation of the product of the antenna response and sky brightness (Section\,\ref{foreground_simulation}). Conceptually, this may be used to model the foreground signal in order to perform a joint fit with the cosmological signal. Independent knowledge of $T_f(\nu,\hat{n})$ and $A(\nu,\hat{n})$ may be gained from interferometric visibilities \citep[i.e.,][]{thompson90}:
\begin{eqnarray}
V_{ij}(u,v,\nu) = \int_{\hat{n}} I(\nu, \ell, m) A(\nu, \ell, m) \, e^{- 2 i \pi (u \ell + vm)} d \ell d m, \nonumber
\end{eqnarray}
where $(i,j)$ represents an antenna pair and $(u,v)$ the separation in units of wavelength, $\nu$ is the observing frequency, $I$ is the specific intensity of the sky, and $(\ell,m)$ are the coordinates on the plane tangent to the sky in the observing direction. 

In practice, reconstructing $I(\nu, \ell, m)$ by inverting this relation is complicated by instrument insensitivity to angular structure much larger than $\lambda/d_{\rm min}$, where $d_{\rm min}$ is the typical minimum separation between antennas. 
This poses a problem for wavelengths $\simgreat3$\,m because the majority of the foreground signal arises on angular scales corresponding to ${d \over \lambda} < 1$, i.e., on very large scales on the sky. Inclusion of autocorrelated power in the inversion \citep[i.e.,][]{stanimirovic02}, as well as forward modeling \citep{bernardi11} of an all-sky template (as assumed in Section~\ref{foreground_simulation}) may mitigate the problem and will be investigated in future work.

We focus here on measurement of the antenna gain pattern using interferometric measurements.  Simulations carried out in Section\,\ref{foreground_simulation} indicated that for intensities comparable to the predicted $\lambda$21\,cm signal, details of the gain pattern impose frequency structure in the observed spectrum more so than the spectrum of foreground brightness itself.  Therefore, we assume in what follows the availability of a foreground template.

\citet{mitchell08} describe an algorithm for gain pattern determination in interferometric observations where visibilities are phased to the positions of calibrators, apparent intensities are measured, and a parameterized antenna gain pattern is fit.  The technique is suitable where the number of sources greatly exceeds the number of gain pattern parameters to be estimated.  Where sensitivity limits the number of calibrators visible instantaneously, measurements for any given object may be made multiple times as it traverses the gain pattern \citep[e.g.,][]{bernardi13}.

We consider application to a model LWA station comprising 256 dual polarization antennas distributed in an area $\sim 100$\,m across (5\,m minimum spacing) plus five dual polarization ``outrigger'' antennas offset $\sim350$\,m from the array center.\footnote{The model array is intended to be generic, borrowing elements of the two LWA sites used in the LEDA effort.  The NM site has a $110\times100$\,m core and five outriggers offset by 213-459\,m from the center.  The Owens Valley site has a 212\,m core and as many outriggers offset in a concentric arc of radius 265\,m.}    Baselines between core and outrigger antennas resolve diffuse Galactic emission and enable isolation of point sources for calibration purposes.  We adopt the 74\,MHz Very Large Array Sky Survey \citep[VLSS,][]{cohen07,helmboldt08} as a point source reference catalog that guides execution of the \citet{mitchell08} algorithm.

Under these assumptions, we can study how well LEDA will be able to constrain the antenna gain pattern by using interferometric measurements. The fractional uncertainty, $\sigma_p$, on each parameter describing  $A(\nu,\ell,m)$ is the inverse of the signal-to-noise ratio per parameter ${\rm SNR}_p$:
\begin{eqnarray}
\sigma_p = \frac{1}{{\rm SNR}_p} & = & \frac{1} {{\rm SNR}_s \sqrt{\frac{N_{\rm chan} N_s}{N_{pd} N_{gp}}}}
\end{eqnarray}
where ${\rm SNR}_s$ is the SNR per source, $N_{\rm chan}$ is the number of independent frequency channels, $N_s$ is the number of detectable sources, $N_{pd}$ is the number of parameters describing $A(\nu,\ell,m)$ \citep[24 for LWA;][]{dowell11}, and $N_{gp}$ is the number of antennas for which the gain patterns are dissimilar one to the other.  The present analysis adopts $N_{gp} = 2$ representing one gain pattern for the geographically isolated outriggers and one mean gain pattern for the closely packed, mutually coupled core antennas.  

The number of detectable sources can be estimated by integrating the 150~MHz differential source counts \citep{hales88}:
\begin{eqnarray}
N_s(S_0) & = & \int_{S_0}^{\infty} \int_{\hat{n'}} \frac{dN}{dS} dS d \hat{n} \nonumber \\
       & = & \int_{S_0}^{\infty} \int_{\Omega}  4000 \left( \frac{S}{{\rm Jy}} \right)^{-2.52} dS d \hat{n},
\label{dn_ds}
\end{eqnarray}
where the angular integral is over the peak of the antenna gain pattern.  (The $-3$ dB points are separated by $\sim90^\circ$ across the band.)  In general, the limit $S_0$ above which sources are detectable is set by either the array sensitivity or the confusion level. Recent observations at 150-200~MHz band have estimated classical confusion noise to be $\sim$200~mJy for $\sim16'$ resolution \citep[]{williams12,bernardi13}. The confusion level $\sigma^{\rm LEDA}_c$ expected for LEDA may be obtained by scaling:
\begin{eqnarray}
\sigma_c^{\rm LEDA} \sim 0.2 \left ( \frac{\nu}{150 {\rm MHz}} \right )^{{\bar \alpha}} \left ( \frac{\theta_{\rm LEDA}}{16'} \right )^2,
\end{eqnarray}
where $\theta_{\rm LEDA}$ is the point spread function expressed in arcmin, and ${\bar \alpha} \sim -0.7$ is the spectral index.  At $\nu = 60$\,MHz, $ \sigma_c^{\rm LEDA} \simeq  20$\,Jy\,beam$^{-1}$ for a baseline of 350\,m.

The  radiometer equation specifies the thermal noise,
\begin{eqnarray}
\sigma^{\rm LEDA} = \frac{{\rm SEFD}}{\sqrt{ {N_a (N_a - 1)\Delta \nu \Delta t }}}
\end{eqnarray}
where the System Equivalent Flux Density (SEFD) for an LWA antenna is 2\,MJy in the middle of the observing band and at mid-range zenith angles \citep[]{ellingson11}, $\Delta \nu$ is the bandwidth, $\Delta t$ the integration time and $N_a = 256$ is the number of correlated antennas (per polarization). For $\Delta \nu = 1$\,MHz and $\Delta t = 1$\,min, $\sigma^{\rm LEDA} \sim 1.4$\,Jy\,beam$^{-1}$, which indicates that the data are confusion limited.   If $S_0$ is $5\times \sigma_c^{\rm LEDA}$, then Equation~\ref{dn_ds} gives $N_s \sim 130$ sources over a 90$^\circ$ field of view.  For $N_{gp} = 2$, $N_{pd}=24$, $N_{\rm chan}=48$ and ${\rm SNR}_s = 5$, we obtain $\sigma_p = 1.3$\%.    

We simulated the impact on foreground modeling of uncertainties $\sigma_p = 1, 3, 5, 8$ and 10\%.  We carried out 100 Monte Carlo realizations where each of the parameters ${\bf a}_i(\nu) \equiv [\alpha_i(\nu), \beta_i(\nu), \gamma_i(\nu), \delta_i(\nu)]$ that describe the beam frequency dependence was perturbed by adding an uncertainty drawn from a Gaussian distribution of zero mean and $\sigma_p$ standard deviation. A set of parameters ${\bf a}_i(\nu)$ was generated for each polynomial order independently in order to assess the impact of uncertainties on different scales of beam frequency dependence. Each simulated spectrum was then fitted by using a $7^{\rm th}$ order polynomial in $\log\nu$ and the residual spectra were averaged together. 

\begin{figure}
\begin{center}
\epsfig{file=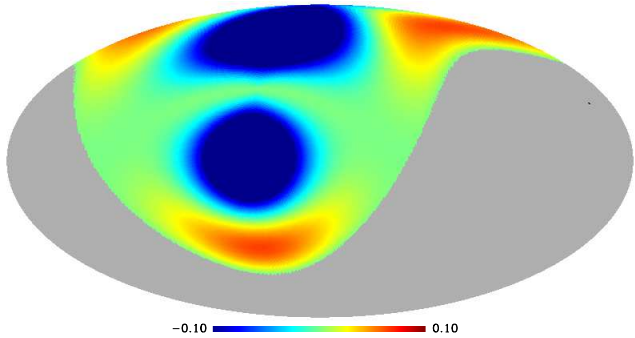,angle = 90,width=8.5cm}
\end{center}
\caption{Difference between the LEDA dipole gain pattern and one realization of the pattern obtained with perturbed parameters (see Section~\ref{leda_approach} for details). The reference frequency is 88~MHz, and both beams are normalized to unity at zenith (see, for instance, Figure~\ref{fig:dipoles}). \label{fig:beam_errors}}
\end{figure}

The beam simulation was repeated by simultaneously varying the ${\bf a}_i$ parameters in order to investigate the impact of their covariance. We found equivalent results for both cases and, therefore, we presented only result for the latter. Figure~\ref{fig:beam_errors} displays one realization of the beam simulations (cf. Figure~\ref{fig:dipoles}). 

\begin{figure}
\begin{center}
\epsfig{file=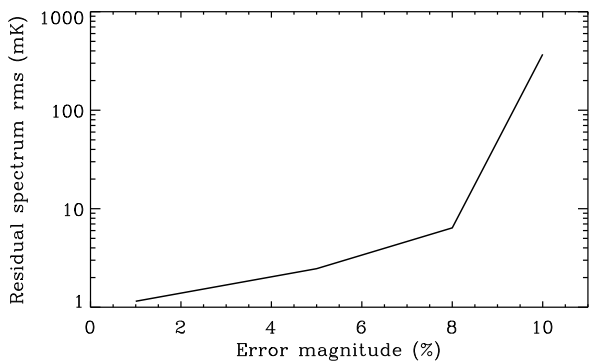,angle = 90,width=8.5cm}
\end{center}
\caption{Residual RMS when fitting with a 7$^{\rm th}$ order polynomial as a function of the fractional uncertainty in antenna gain pattern parameters. We estimate that LEDA can constrain the dipole gain pattern parameters to 1.3\% using interferometric data, which would result in negligible increase in residuals over the case where these parameters are perfectly known.\label{fig:calibration_errors}}
\end{figure}

We found that increasing the error magnitude leads to higher RMS residuals assuming that the polynomial order of the fitting function is not changed, i.e., mimicking incomplete knowledge of the gain pattern (Figure~\ref{fig:calibration_errors}). For errors $<5\%$ the residual RMS remains at the few mK level and well below the anticipated $\sim 100$\,mK signal.  
From this we infer that, in principle, the LEDA gain pattern could be constrained with sufficient accuracy so as not to motivate adoption of a higher order representation than what is already required (Figure\,\ref{fig:residual_fit}, \ref{fig:residual_fit3}).

\section{Fisher analysis}
\label{sec::fisher}

\begin{figure}
{\epsfig{file=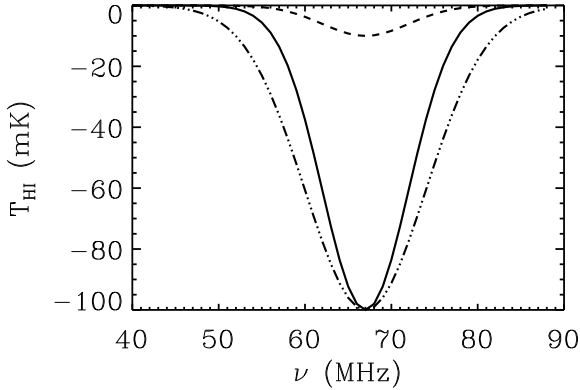, width=8.5cm}}
\caption{$\lambda$21\,cm line profiles corresponding to the models listed in Table~\ref{HI_models}: the solid line corresponds to model A, the triple dot-dashed line to model B and the dashed line to model C.\label{fig:hi_profiles}}
\end{figure}

Thus far, we have shown that the residuals after subtracting a low order polynomial in $\log{\nu}$ are less than the physical $\lambda$21\,cm signal amplitude, which suggests that the signal is detectable.  However, foreground subtraction may also remove signal and we have not formally shown that the anticipated $\lambda$21\,cm signal is detectable after fitting out an, e.g., $7^{\rm th}$ order polynomial.  We again specialize this section to the case of LEDA.  To estimate how precisely LEDA can constrain the HI absorption trough, we use the Fisher matrix formalism \citep[i.e.,][]{tegmark97,eisenstein99}, which provides analytic formula for errors in the limit of a Gaussian likelihood function.  This formalism requires a model for the foregrounds, signal, and noise.
We adopt the foreground model developed in Section~\ref{foreground_simulation}:
\begin{eqnarray}
T_{\rm f}(\nu) = e^{\sum_{n=0}^N c_n (\log{\nu_1})^n},
\label{fish_for_model} 
\end{eqnarray}
where the $c_n$ coefficients are the best fit derived by the foreground simulations and we use $N=7-8$. 
For the $\lambda$21\,cm signal, we adopt a Gaussian: 
\begin{eqnarray}
T_{\rm HI}(\nu) = A_{\rm HI} \, e^{-\frac{(\nu - \nu_{\rm HI})^2}{2 \sigma^2_{\rm HI}}},
\end{eqnarray}
where $A_{\rm HI}$, $\nu_{\rm HI}$ and $\sigma_{\rm HI}$ are the peak amplitude, peak frequency and width of the HI signal respectively.   We use three Gaussian profiles (Table~\ref{HI_models} and Figure~\ref{fig:hi_profiles}) to model the HI line and to coarsely sample the range of the theoretical predictions \citep{pritchard08, pritchard10}.  These Gaussian models differ somewhat from the approach taken in \cite{pritchard10} and \cite{harker12}, that modeled the cosmological $\lambda$21\,cm signal by its turning points, i.e., where the derivative of the signal with respect to the frequency is zero. This approach is simply not possible for LEDA as only one turning point tends to fall within the LEDA observing band. For the same reason, the models investigated here span a narrower range of frequencies than is shown in Figure~\ref{fig:global}.\\

\begin{table}	
\begin{center}
\caption[]{$\lambda$21\,cm line profile models used in our Fisher error estimates.} 
\label{HI_models}
\begin{tabular}{l l l l}        
\hline\hline 
Model label				& $A_{\rm HI}$ (mK)	& $\nu_{\rm HI}$ (MHz)	& $\sigma_{\rm HI}$ (MHz) \\
\hline 
A				& -100			& 67			& 5	\\				
B				& -100			& 67			& 7	\\
C				& -10			& 67			& 5	\\
\hline
\end{tabular}
\end{center}
\end{table}

The Fisher matrix is defined as
\begin{eqnarray}
F_{ij} = \frac{1}{2} \rm{Tr} [C^{-1}C_{,i}C^{-1}C_{,j} + C^{-1}(\mu_{,i}\mu^{\rm T}_{,j} + \mu_{,i}\mu^{\rm T}_{,j})],
\end{eqnarray}
where $C$ is the covariance matrix between the frequency channels, $\mu \equiv T_{\rm sky}(\nu) = T_{\rm f}(\nu) + T_{\rm HI}(\nu)$, and commas represent derivatives with respect to parameter $i$. Assuming the signal is measured in $N_{\rm chan}$ uncorrelated frequency channels, the covariance matrix takes the following form:
\begin{eqnarray}
C_{n,m} = \delta_{n,m} \sigma^2_n = \delta_{n,m} \frac{T^2_{\rm sky}(\nu_n)}{\Delta \nu \Delta t},
\end{eqnarray}
and the Fisher matrix becomes \citep{pritchard10}:
\begin{eqnarray}
F_{ij} & = & \sum^{N_{\rm chan}}_{n=1} \left[2 + \frac{ \Delta \nu \Delta t}{T^2_{\rm sky}(\nu_n)} \right] \frac{dT_{\rm sky}(\nu_n)}{dp_i} \frac{dT_{\rm sky}(\nu_n)}{dp_j},
\label{fisher_eq}
\end{eqnarray}
where ${\bf p} \equiv (c_0,...,c_N,A_{\rm HI},\nu_{\rm HI},\sigma_{\rm HI})$ and we have assumed $\Delta \nu = 1$~MHz, $N_{\rm chan} = 48$ and $\Delta t = 400$~h. 
As LEDA has deployed four antennas at each site for the global signal measurement, this is equivalent to an integration time of 100~h per antenna.

\begin{figure*}
{\epsfig{file=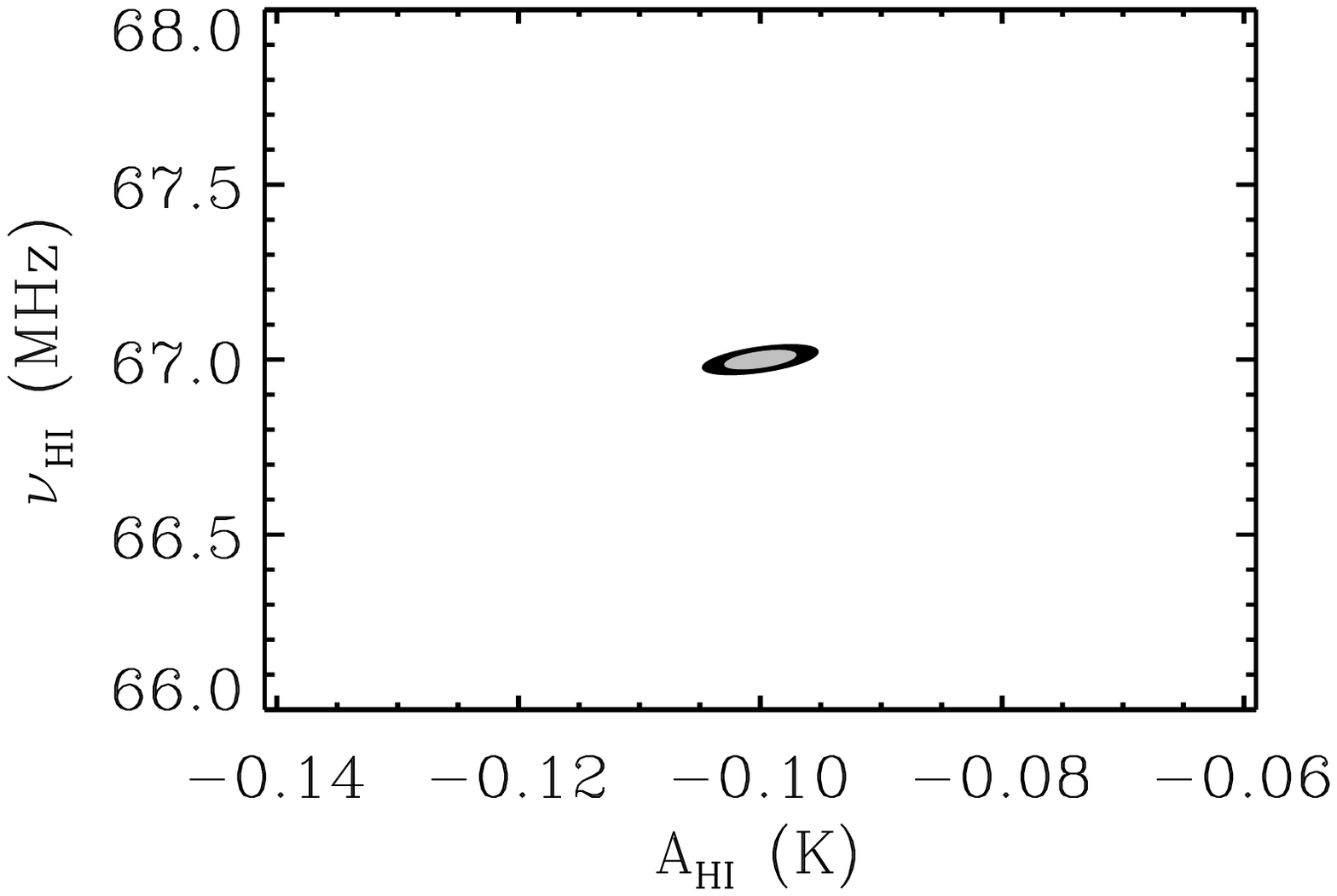, width=6cm}}
{\epsfig{file=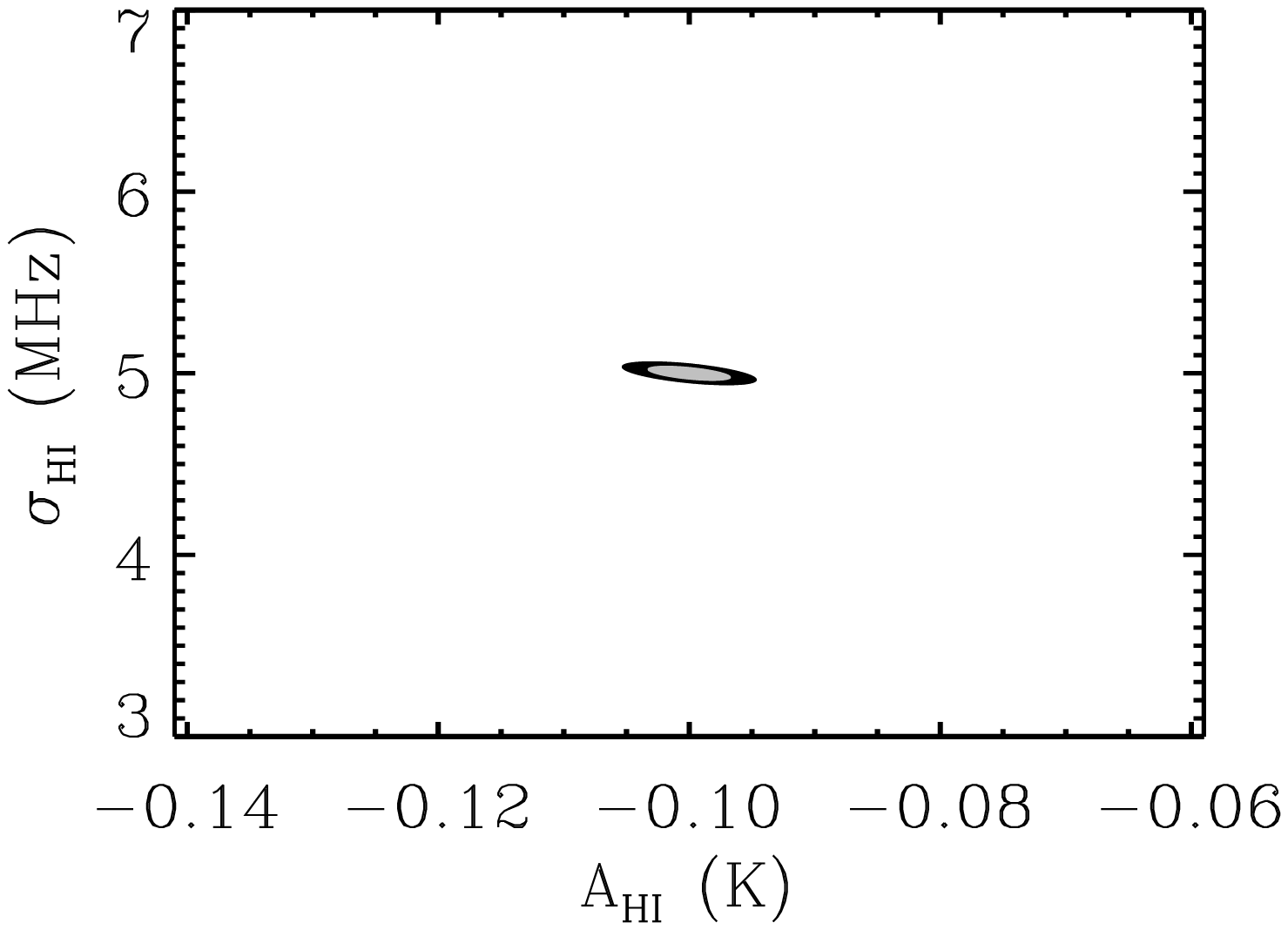, width=6cm}}
{\epsfig{file=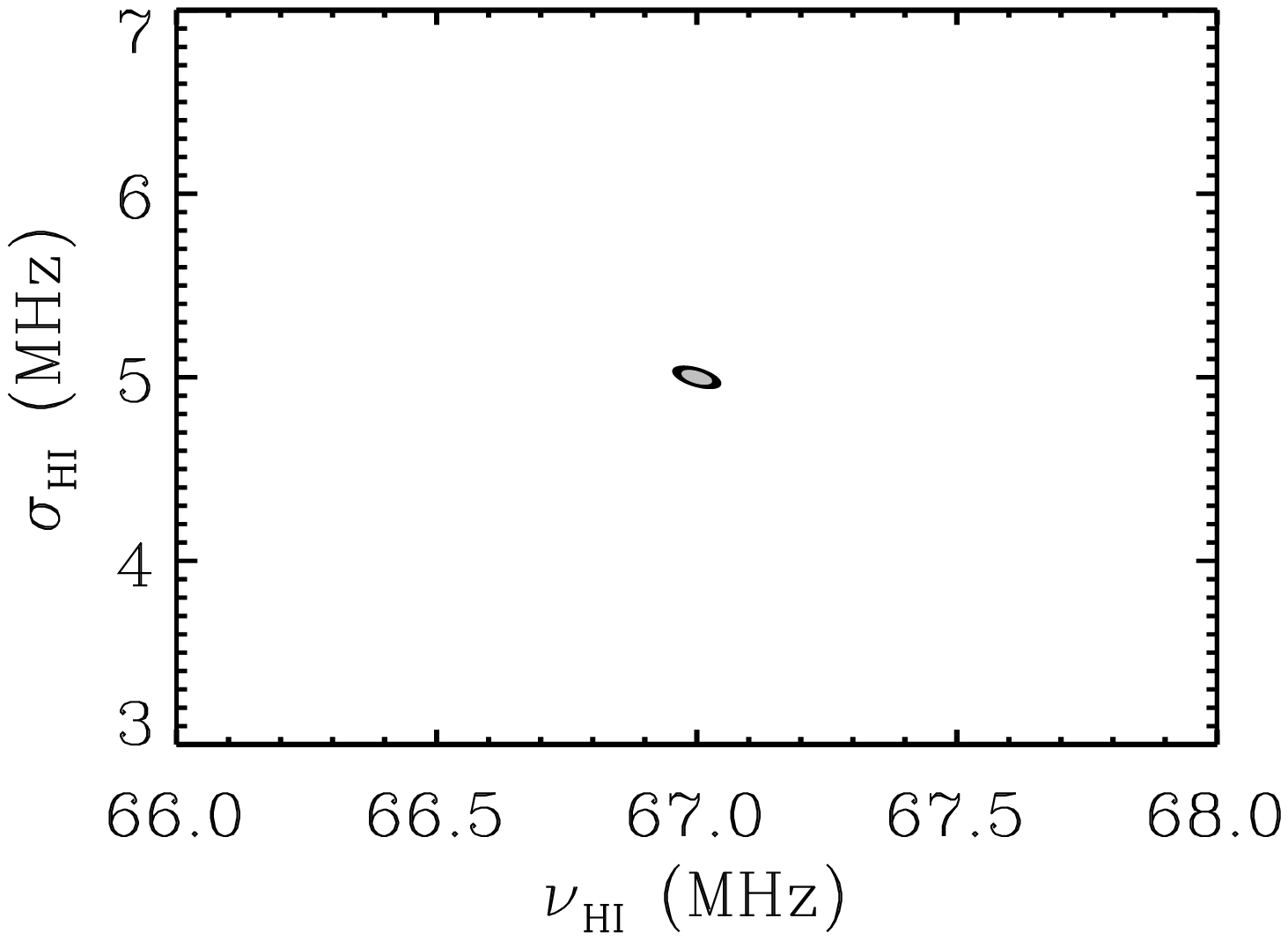, width=6cm}}
{\epsfig{file=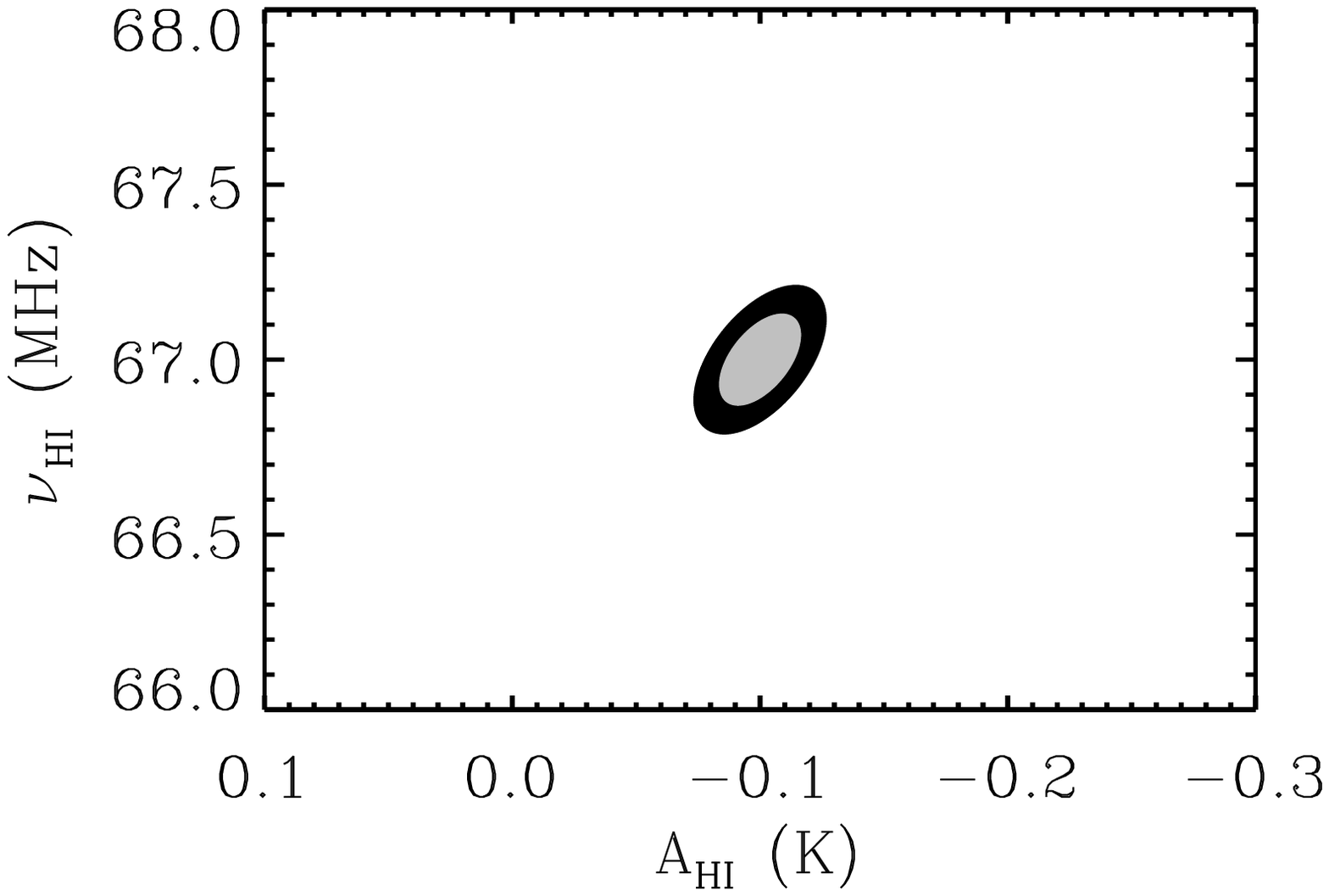, width=6cm}}
{\epsfig{file=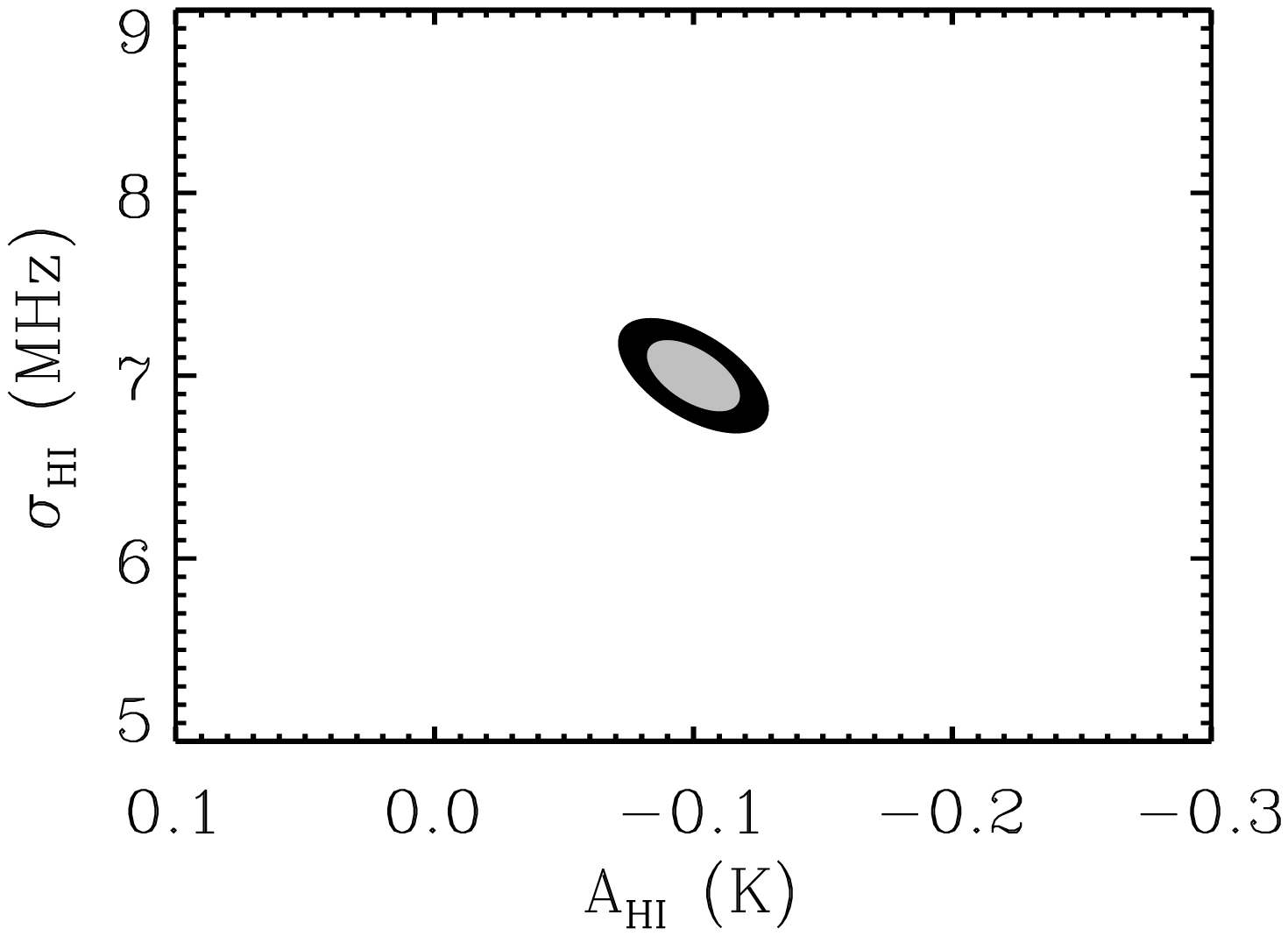, width=6cm}}
{\epsfig{file=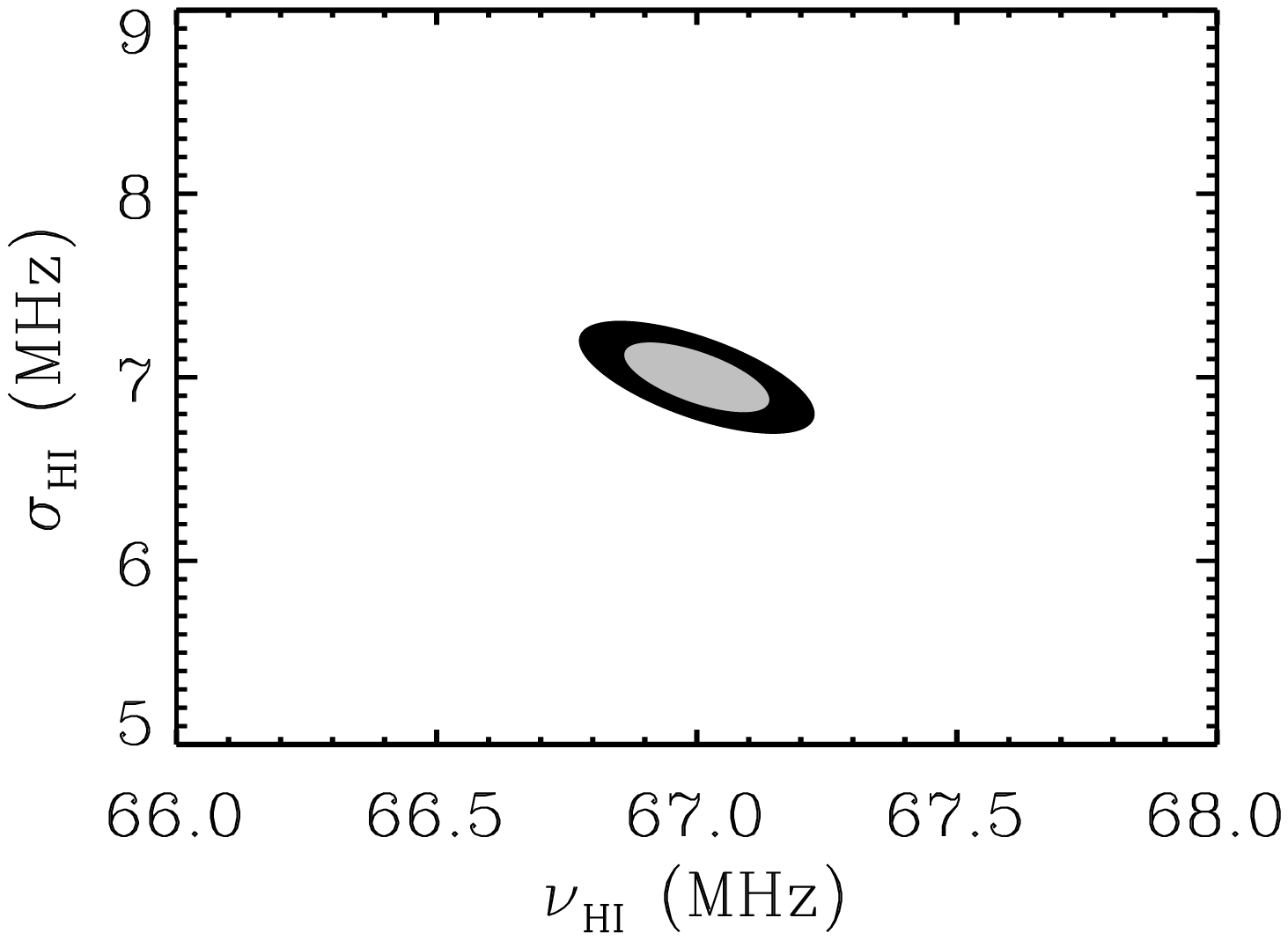, width=6cm}}
{\epsfig{file=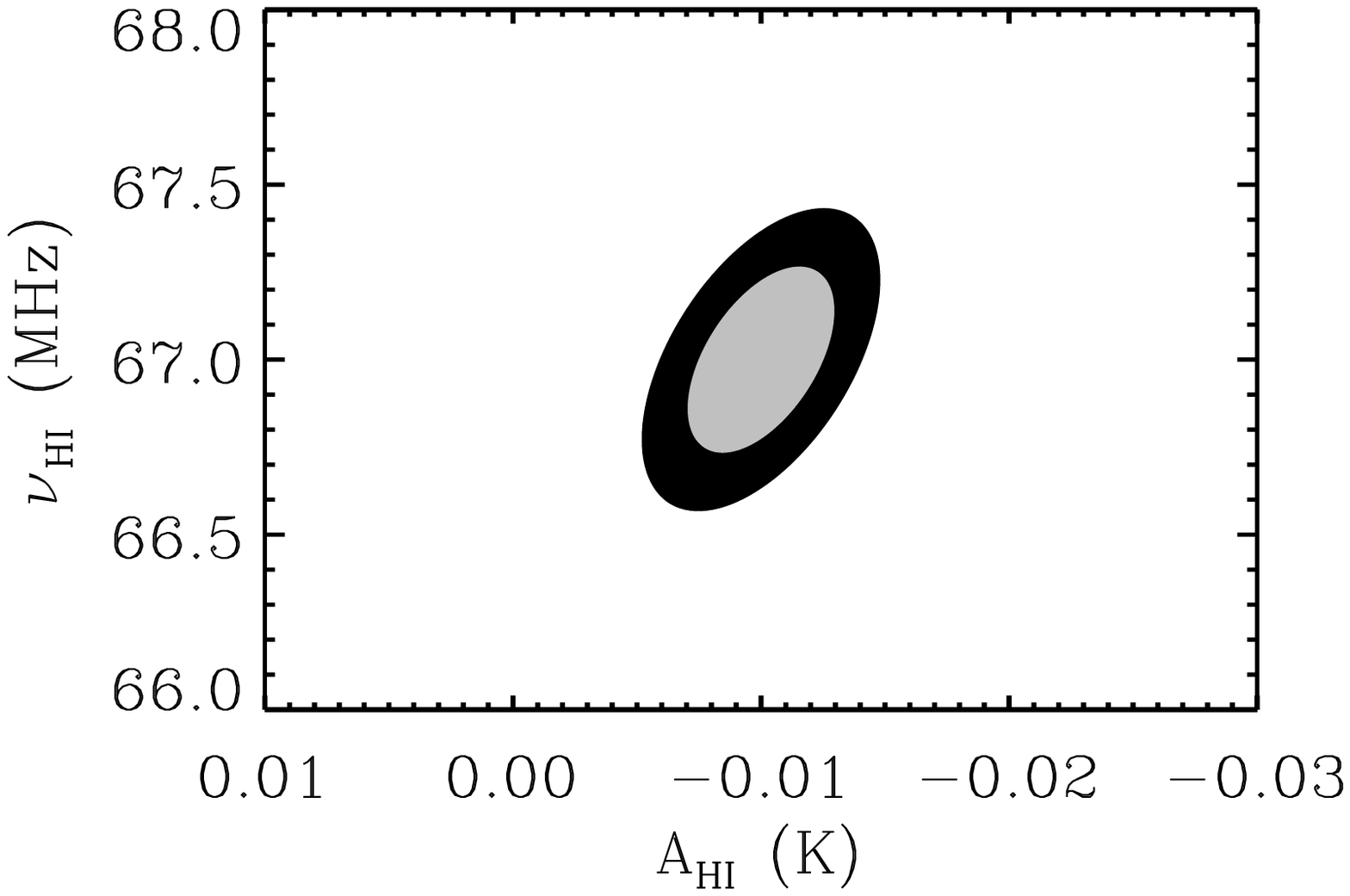, width=6cm}}
{\epsfig{file=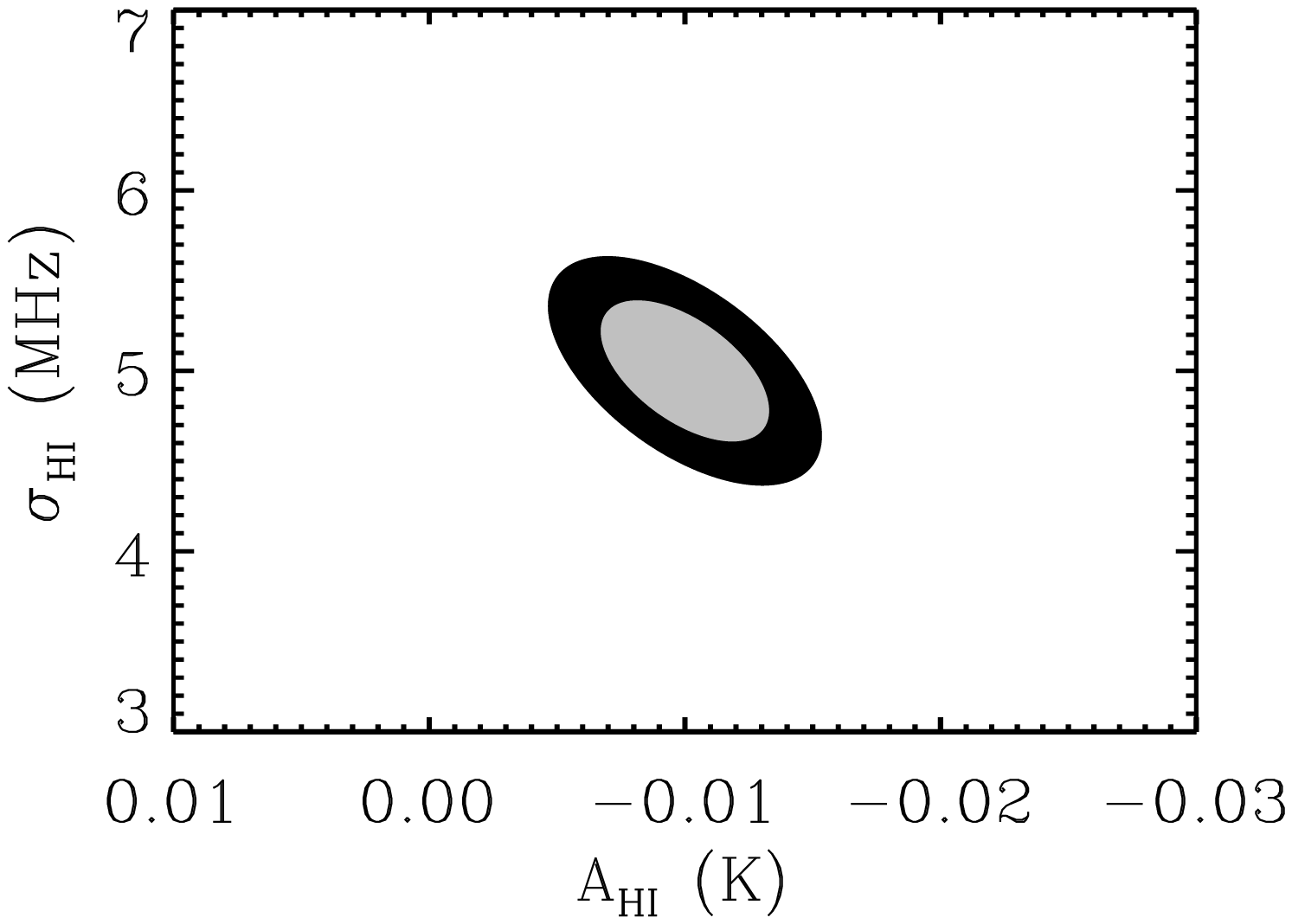, width=6cm}}
{\epsfig{file=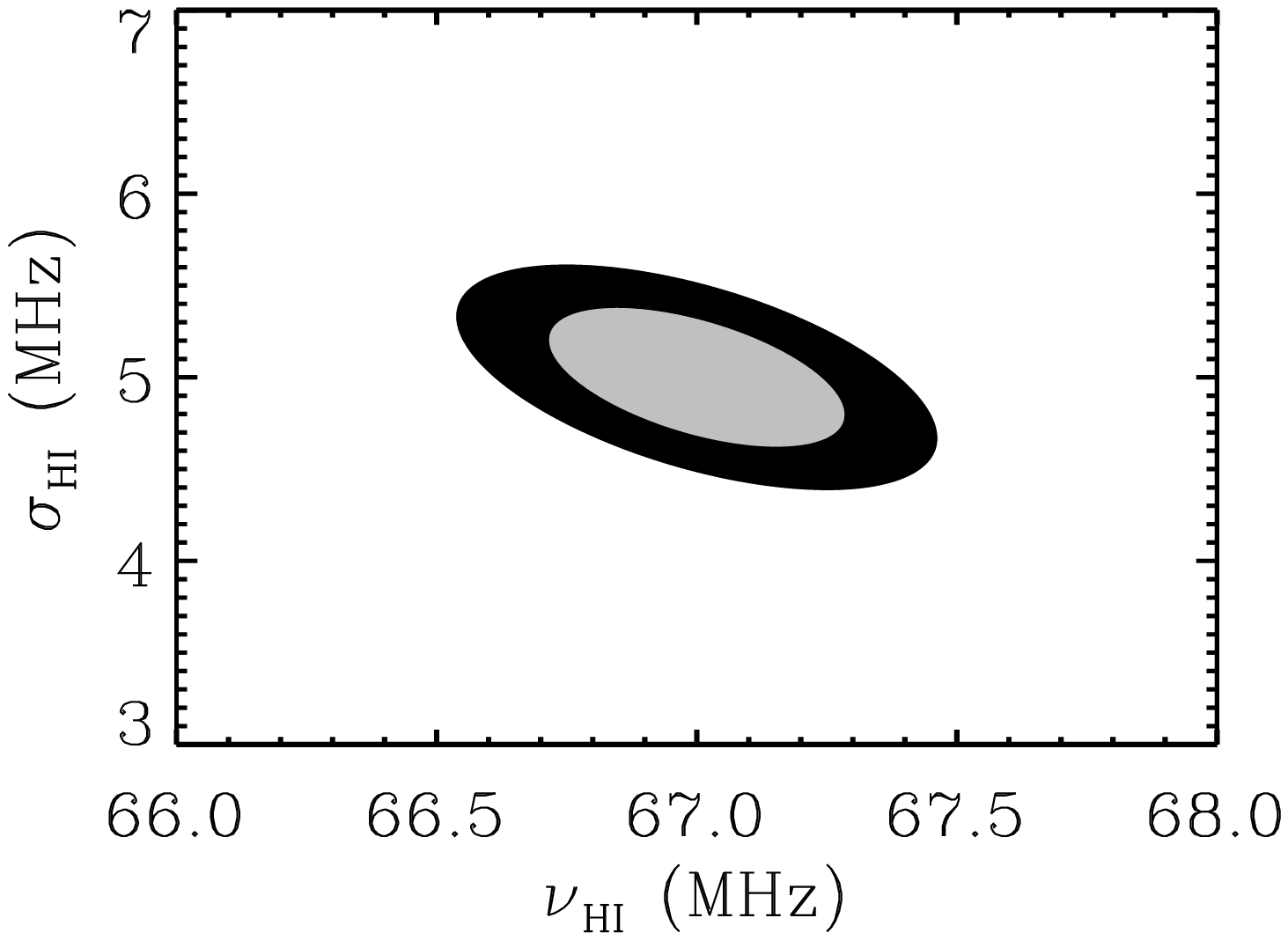, width=6cm}}
\caption{The $68\%$ (gray) and $95\%$ (black) confidence regions for the HI models A (top row), B (second row) and C (third row), assuming the fiducial 7$^{\rm th}$ order polynomial model for the foregrounds.  All the models are detected with high confidence.\label{fig:fisher}}
\end{figure*}

\begin{figure*}
{\epsfig{file=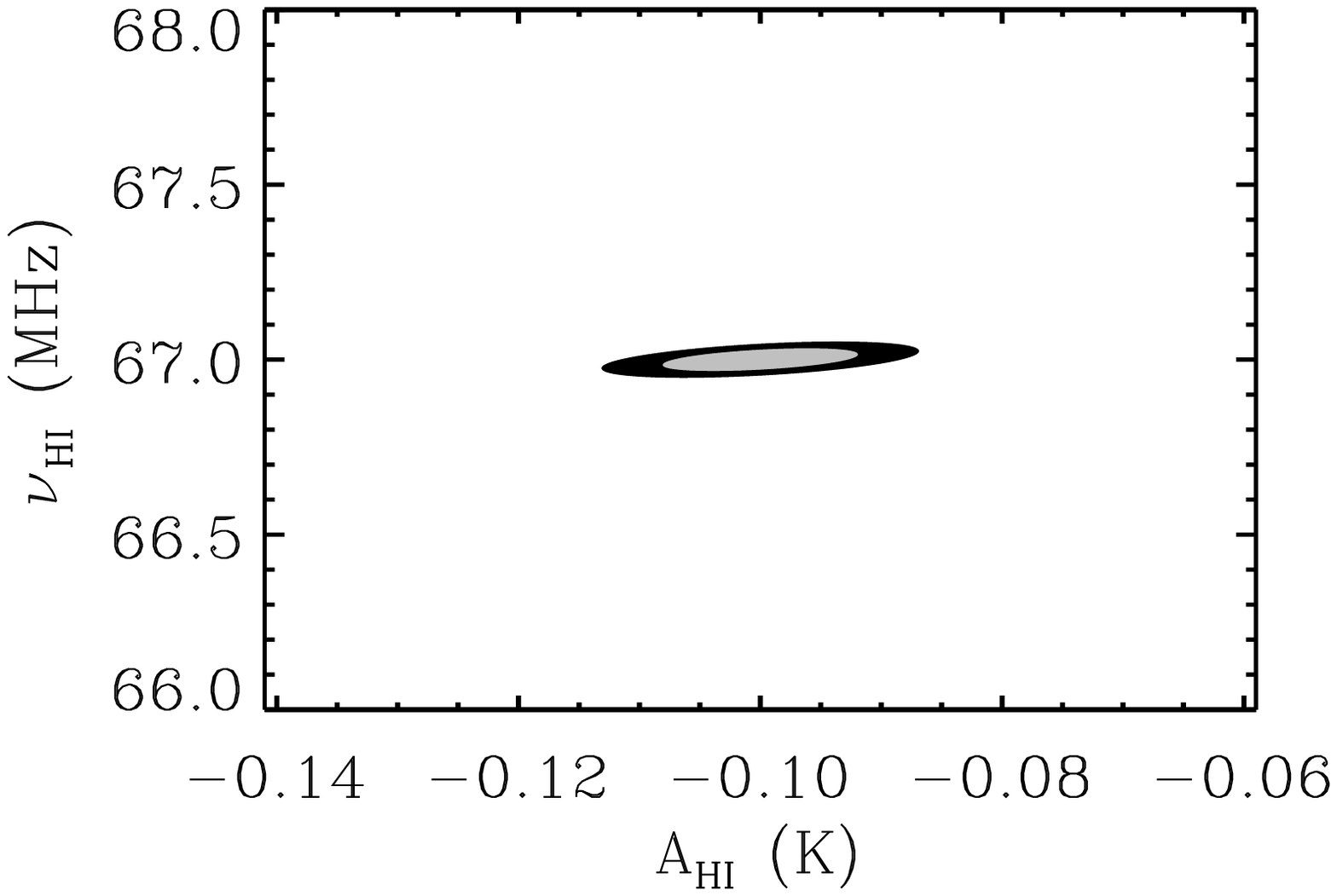, width=6cm}}
{\epsfig{file=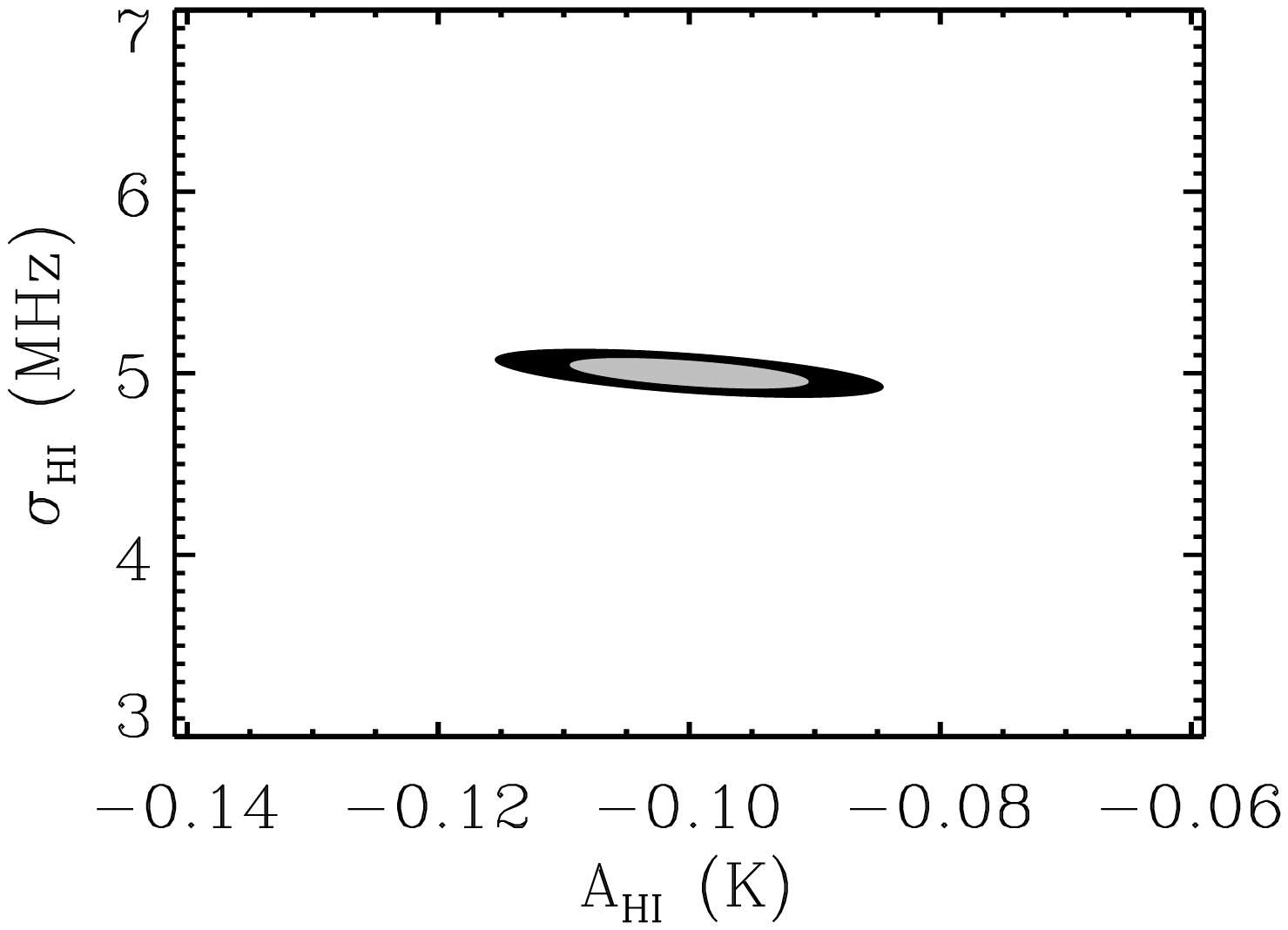, width=6cm}}
{\epsfig{file=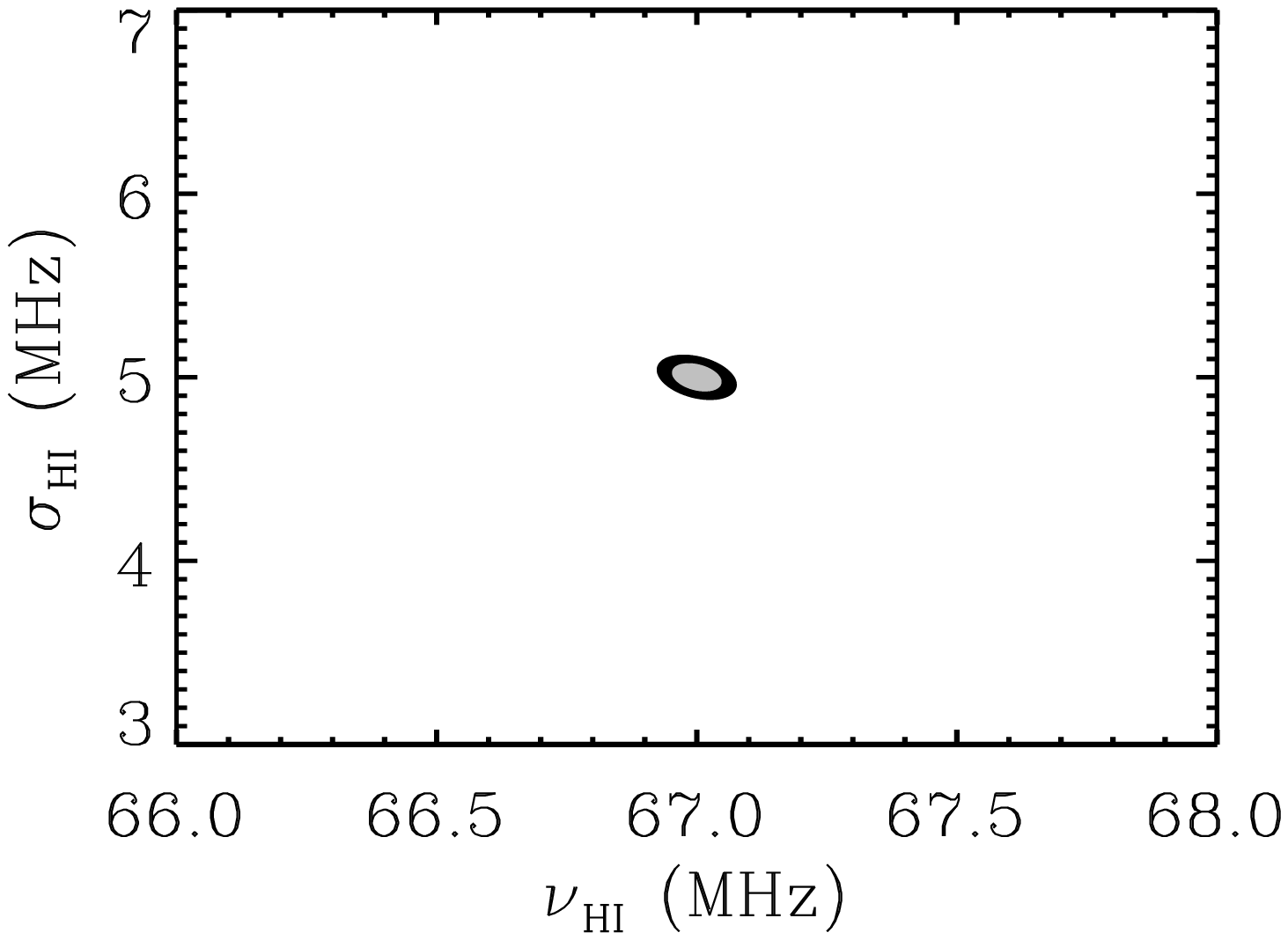, width=6cm}}
{\epsfig{file=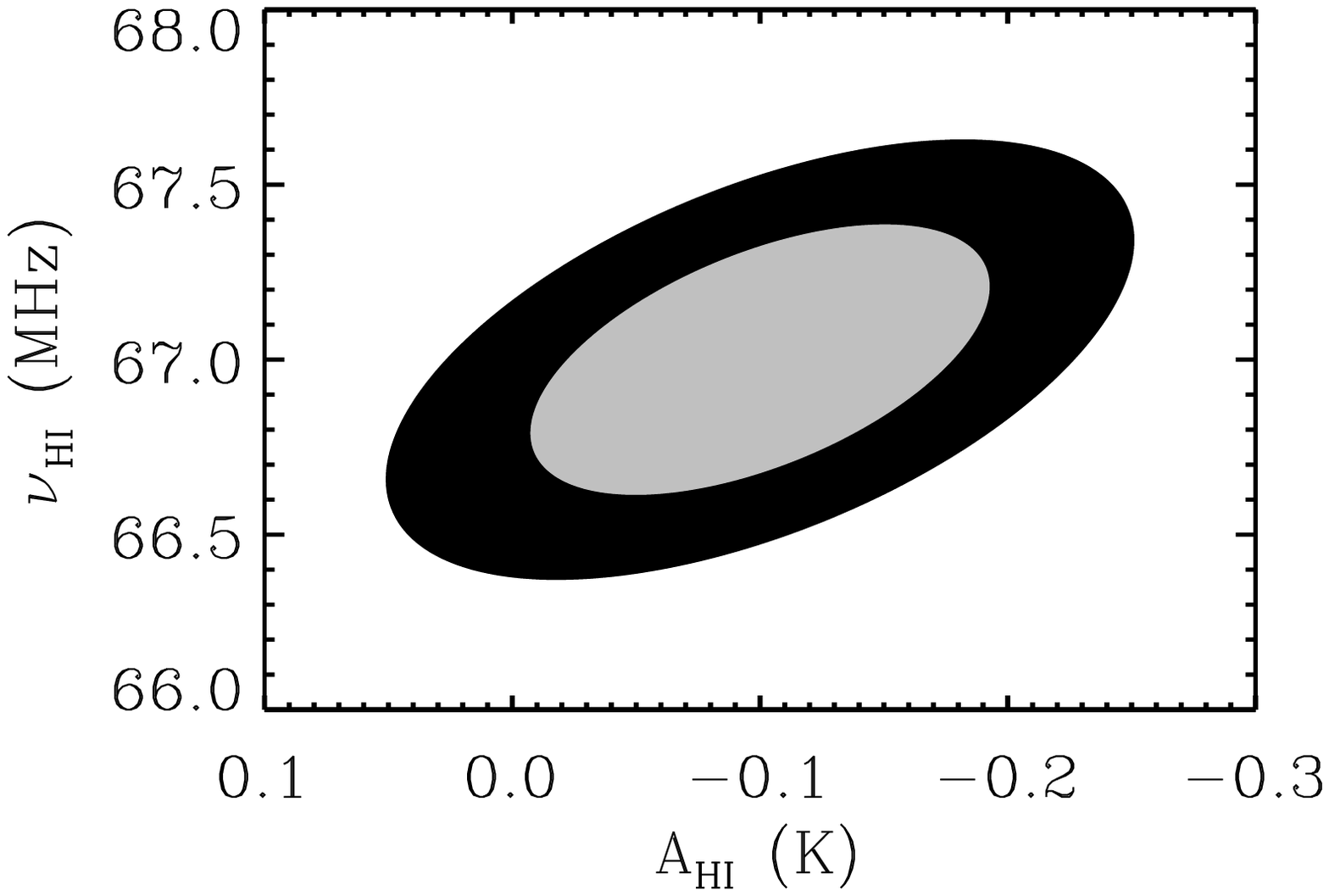, width=6cm}}
{\epsfig{file=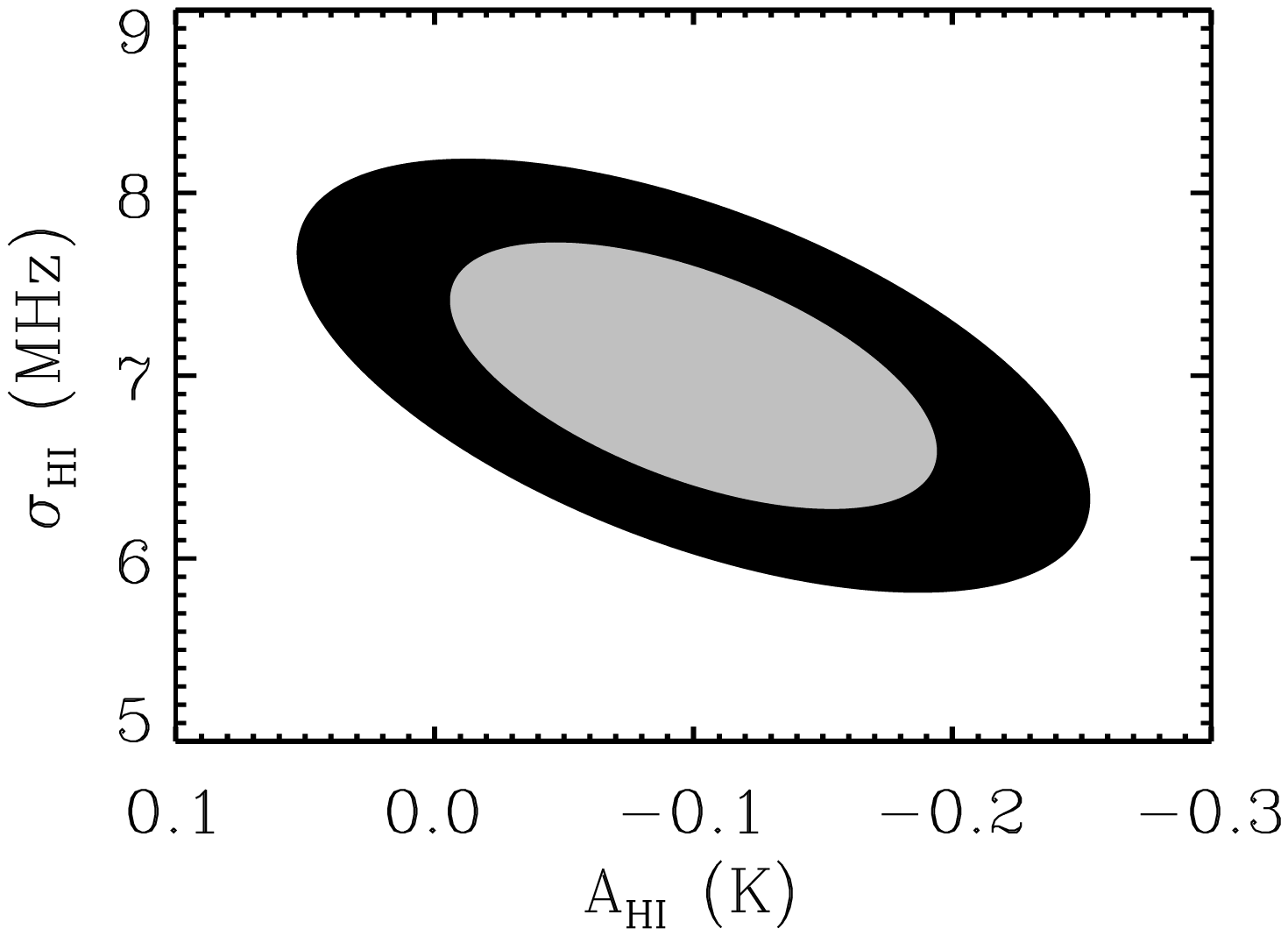, width=6cm}}
{\epsfig{file=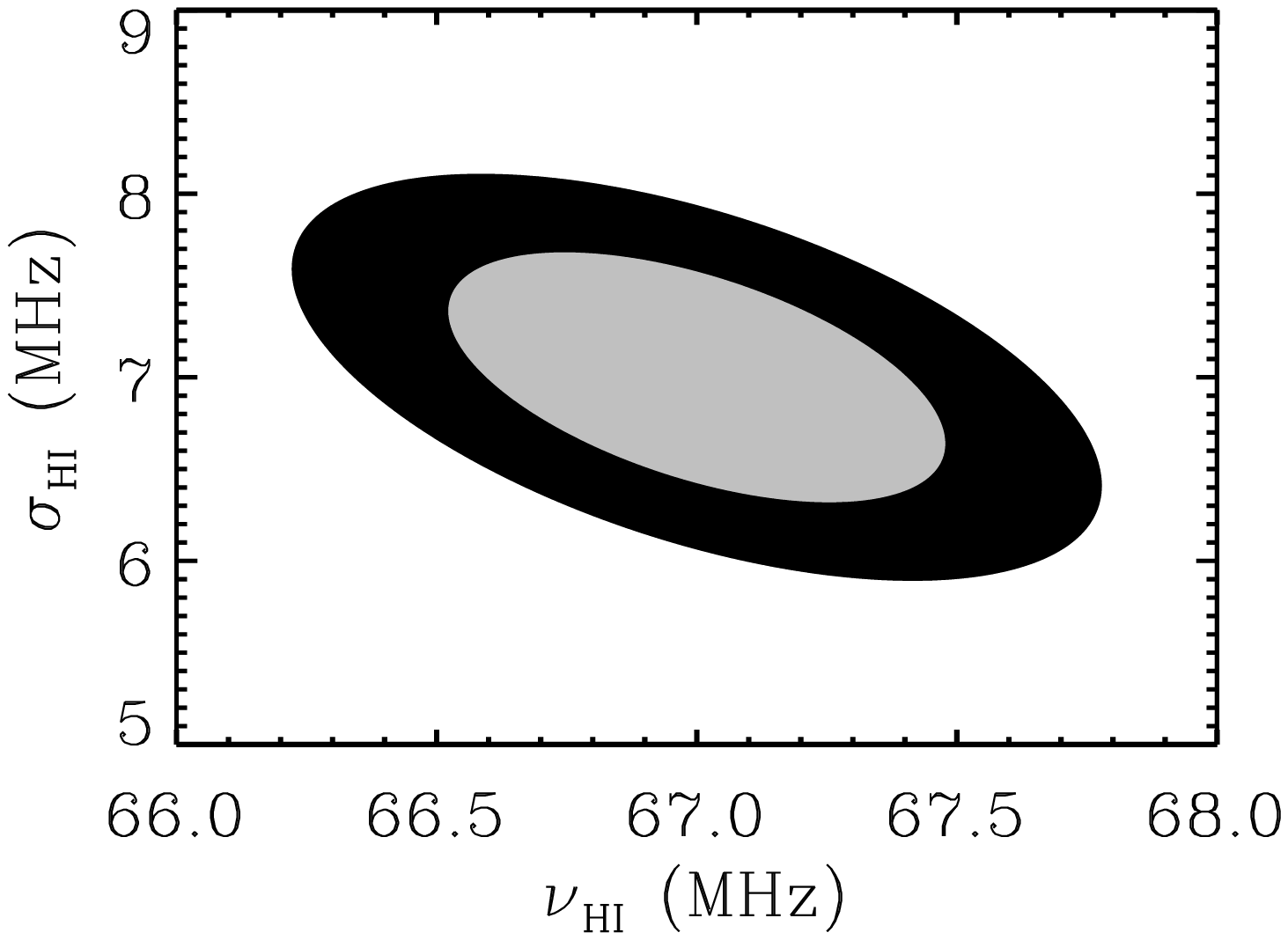, width=6cm}}
{\epsfig{file=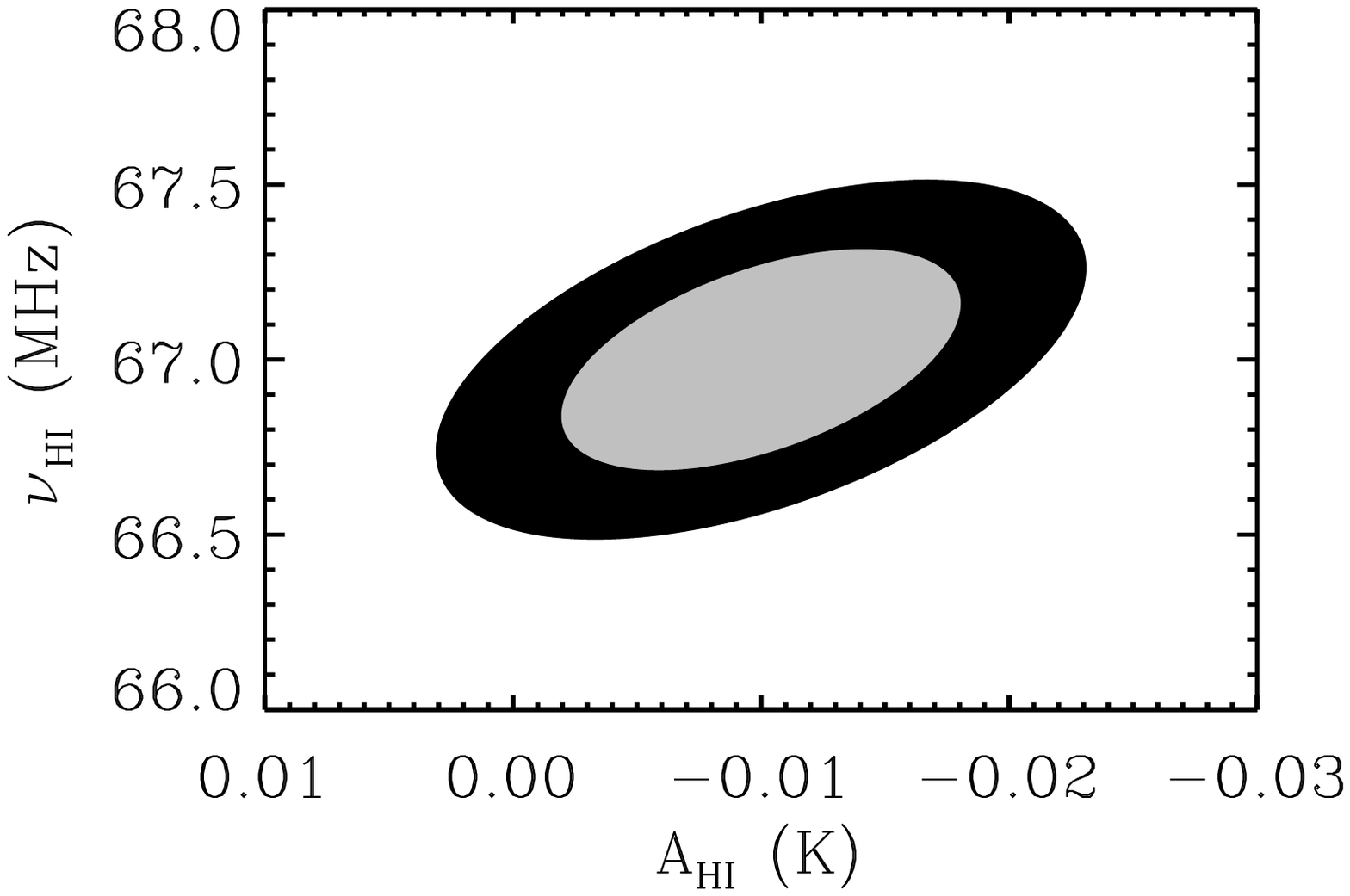, width=6cm}}
{\epsfig{file=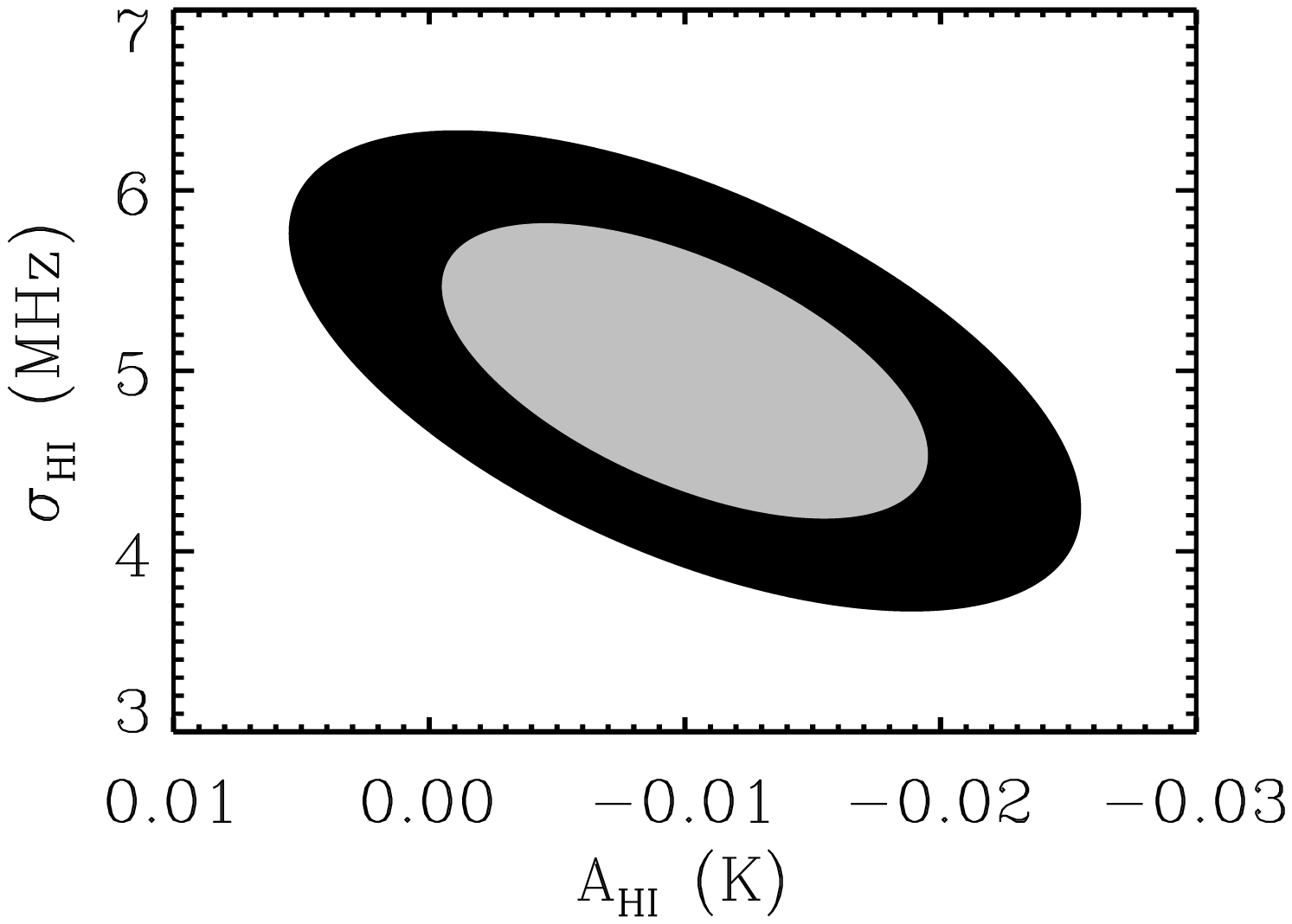, width=6cm}}
{\epsfig{file=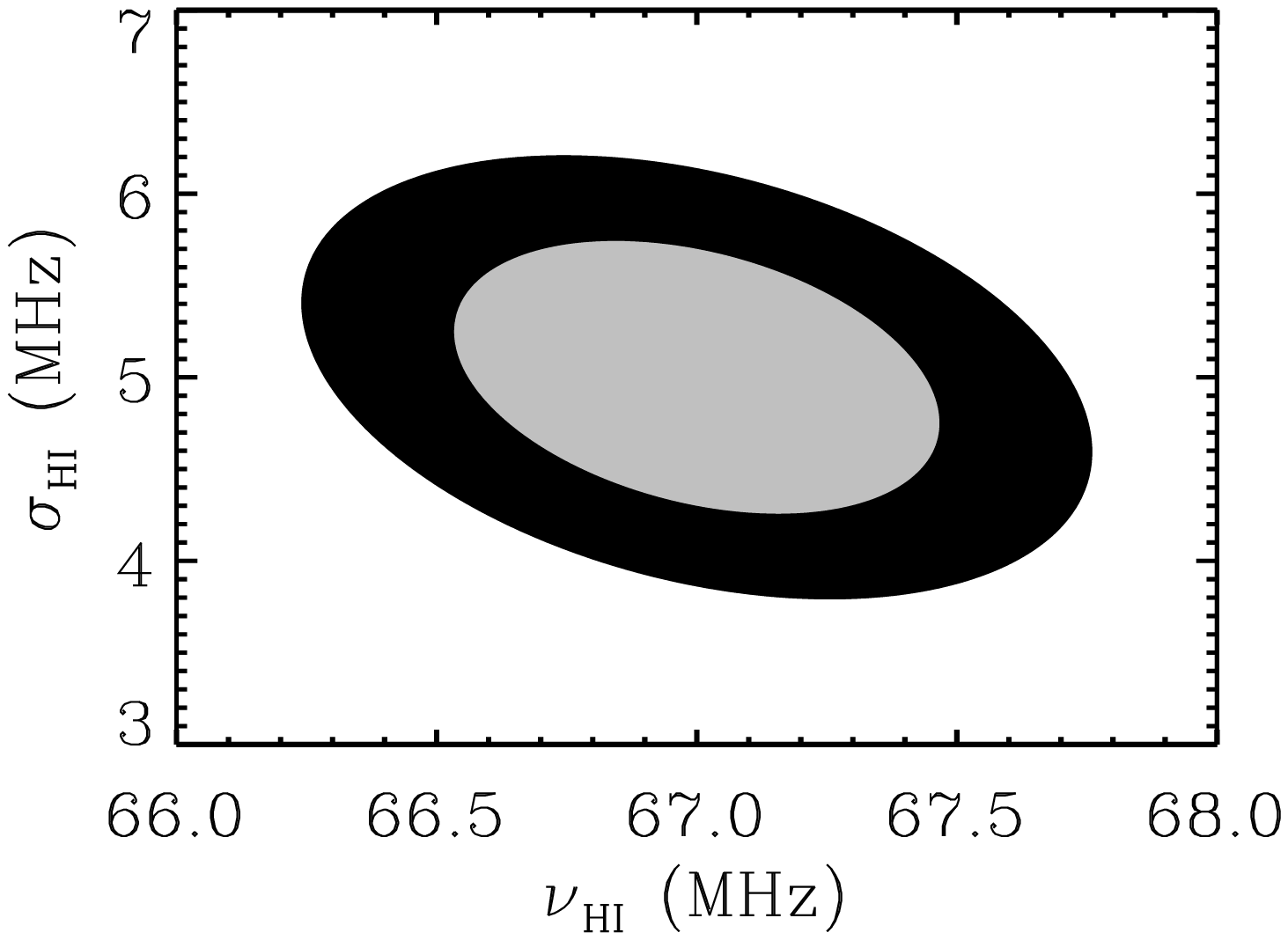, width=6cm}}
\caption{Same as Figure~\ref{fig:fisher}, but for an 8$^{\rm th}$ order polynomial model for the foregrounds.  The bias is smaller in this model than for the 7$^{\rm th}$ order, but the errors are considerably larger such that only Model A is detected with a high confidence level. \label{fig:fisher_8th}}
\end{figure*}

Figure \ref{fig:fisher} shows our Fisher estimates for the errors on the three parameters, assuming the foregrounds are fitted with $N=7$.  The Fisher estimate for the standard deviation on parameter $i$ is $\sqrt{[F^{-1}]_{ii}}$.  We find that in this case all the parameters in models A, B, and C can be detected at 95\% confidence level.  The trough position, $\nu_{\rm HI}$, is the best constrained parameter of the three. The HI amplitude is the parameter which is least constrained, with a percent error of $\sim$40\% for model B. The uncertainties on the width are larger for wider Gaussians, as expected.  Figure \ref{fig:fisher_8th} instead shows the case with $N=8$.  In this case, only model A is detected at 95\% confidence level, which stresses the importance of reducing the order required to fit the polynomial.  We now discuss what value of $N$ to choose.

The residuals in the foreground model that remain after fitting at $N^{\rm th}$ order bias the inference of the $\lambda$21\,cm signal model parameters.  When this bias is larger than the statistical error, it is significant and means that a higher order polynomial is required.  In the Fisher Matrix formalism, one can estimate the bias on a parameter in units of the variance on the parameter $[{F}^{-1}]_{ii}$:
\begin{eqnarray}
b_i = \frac{1}{[{F}^{-1}]_{ii}} \sum_{j=1}^{N_{\rm par}} F_{ij}^{-1} \sum^{N_{\rm chan}}_{n=1} \frac{\Delta \nu \Delta t}{T^2_{\rm sky}(\nu_n)} \frac{d T_{\rm sky}(\nu_n)}{d \lambda_j}R(\nu_n)
\label{bias_eq}
\end{eqnarray}
where $N_{\rm par}$ is the number of parameters to be estimated, $\lambda_j$ is the $j^{\rm th}$ parameter and $R$ is the residual temperature, which we compute from the difference between our full model for the signal with our best fit using equation~(\ref{fish_for_model}). This formula is most appropriate when the bias is comparable or smaller than the error, but it can be used to provide a sense for when this condition should hold.

\begin{figure}
{\epsfig{file=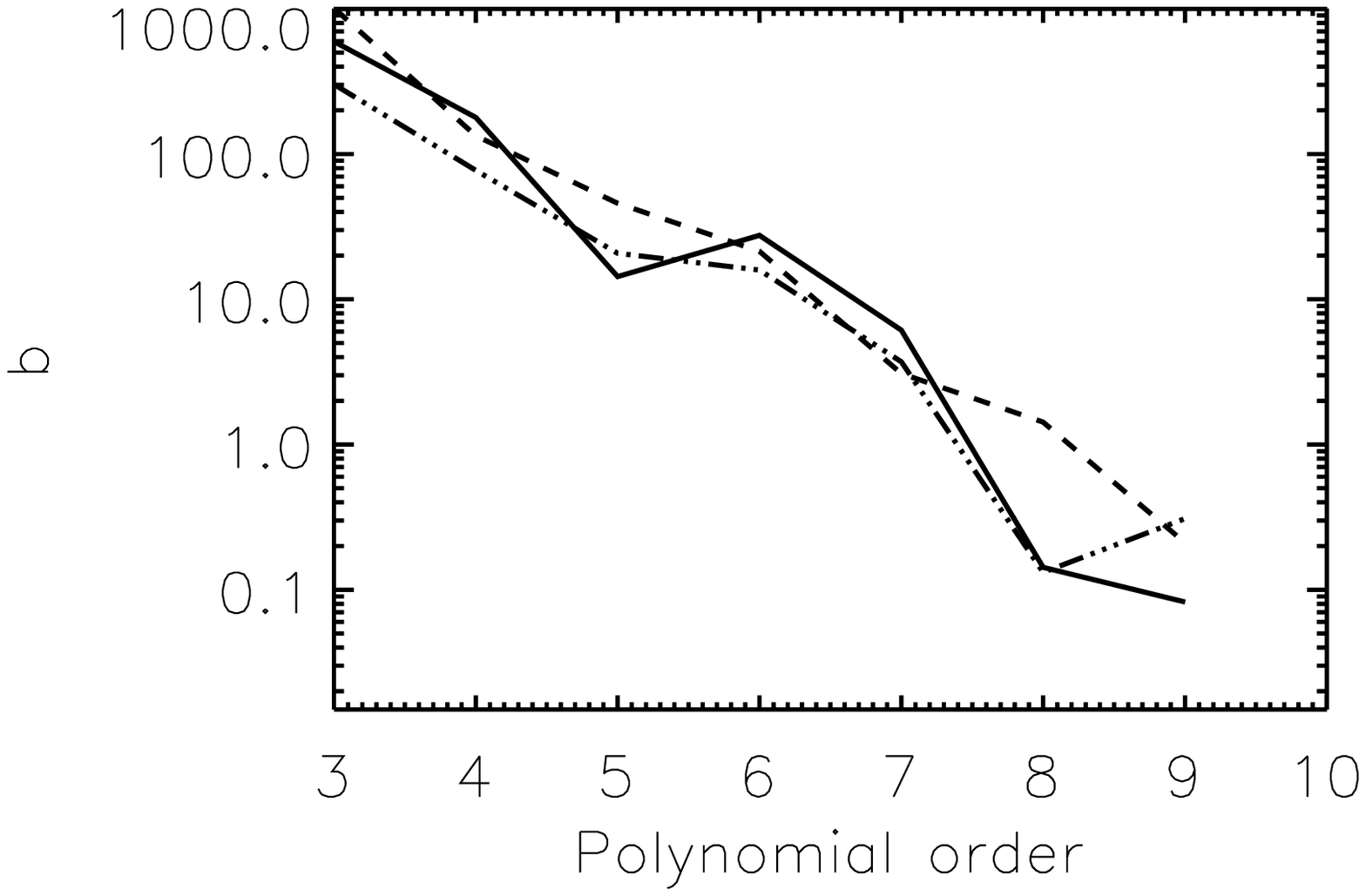,width=8.5cm}}
\caption{Bias in the HI parameters in units of the standard deviation for our mock observation in Model A. It is plotted as a function of the polynomial order used to model the foregrounds.  The solid, dashed and dot-dashed lines show respectively the bias in the amplitude ($A_{\rm HI}$), the centroid ($\nu_{\rm HI}$), and the Gaussian width ($\sigma_{\rm HI}$) of the $\lambda$21\,cm absorption trough.  Note that this function should not be strictly monotonically decreasing for one foreground realization.  \label{fig:bias}}
\end{figure}

Figure~\ref{fig:bias} plots the absolute value of the bias of the three parameters in Model A and as a function of polynomial order.  Again, this needs to be approximately less than unity to consider the parameter estimate unbiased.  We find that $|b_i|$ is a strong, decreasing function of the polynomial order used to fit the foregrounds.  The bias in the model parameters is several times larger than the variance for $N=7$ and smaller than the variance for $N=8$, justifying our previous choices for $N$. 


\section{Conclusions}
\label{sec::conclusions}

This paper investigated the impact on global $\lambda$21\,cm signal measurements of spectral structure that is either intrinsic to the foregrounds or induced by the antenna response. We focused on the measurement of the HI signal from the dark age, focusing on  the LEDA instrument, but our results generalize to other sky-averaged $\lambda$21\,cm experiments. We carried out realistic simulated observations, including a variety of physically-motivated $\lambda$21\,cm signals, foreground models, and antenna gain patterns.

First, we considered more physical models for the spectral structure of the foregrounds than done previously.  Prior studies had primarily parametrized the foregrounds with a low order polynomial in $\log T_b - \log \nu$.  We found that the intrinsic spectral shape of optically-thin synchrotron emission -- the dominant foreground -- likely can be removed with a $\sim 4^{\rm th}$ order polynomial to $<10$~mK, as is required to isolate the signal.  Even in the most pathological case imaginable of a mono-energetic population of synchrotron electrons, we found that the foregrounds could be subtracted with a $\sim 6^{\rm th}$ order polynomial.  We argued that including the additional emission from point sources, free-free emission, and self-absorption requires a modest increase in polynomial order over that required in the optically-thin synchrotron case.  

The primary result of this paper concerns the coupling between the antenna response to the foregrounds.  We found that the coupling between the foregrounds and the (inevitable) spatially and spectrally-dependent antenna gain pattern generates spectral structure that requires additional orders to subtract the foregrounds. 
For LEDA, this induced structure can still be adequately fit with a $7^{\rm th}$ order polynomial in $\log{\nu}$ and a large variety of HI models can still be measured. We found, however, that the subtraction of a $7^{\rm th}$ order polynomial can still leave a non negligible bias in the estimate of the HI parameters. The inclusion of an $8^{\rm th}$ order polynomial in the foreground modeling may preclude the detection of the faintest HI models but leads to an unbiased estimate of the parameters. {As LEDA observes in the 40-90~MHz band, our conclusions are limited to $\lambda$21cm absorption lines narrower than this band. We would expect worse constraints for wider $\lambda$21cm models.

In general, our results suggest that the commonplace approach of assuming \emph{all} frequency structure can be removed with a low order (i.e., $3^{\rm rd}$) function is too optimistic.
The combination of interferometric array and single dipole observations can in principle enable the measurement of the antenna parameters and mitigate antenna gain pattern uncertainties as a source of spectral structure.  We simulated the impact of statistical errors on the antenna gain pattern and found that interferometric measurements should constrain sufficiently the antenna gain pattern so as not to compromise the measurement of the cosmological signal.

This study ignored two effects: (1) the additional frequency structure generated at the stage of signal acquisition that is independent of the antenna gain pattern (i.e., signal reflection due to cable mismatch) and (2) ionospheric refraction and atmospheric absorption.  Accounting for the structure in frequency of the sky-averaged signal resulting from the former effect requires a more complete model of the instrument \citep[i.e.,][]{bowman08,rogers12} that will be the focus of future work. Regarding the latter, the ionosphere can have a twofold effect. First, it can affect the measurement of the antenna primary beam \citep[i.e.,][]{tasse13}. As the ionosphere is expected to vary on time, frequency and spatial scales that are very different than the antenna beam, the problem can effectively be decoupled and ionospheric effects corrected by providing a list of point sources that can be used to measure ionospheric offsets \citep[]{mitchell08}. Second, chromatic effects (owing to refraction in the upper atmosphere) might induce curvature in the spectrum of the global sky signal \citep{vedantham13}, and compromise the logarithmic--space foreground removal method (although not necessarily for the $7-8^{\rm th}$ order polynomials in $\log{\nu}$ considered here).  However, \citet{vedantham13} also showed that ionospheric effects can be understood with simple physical principles, which suggests that they can be modeled prior to fitting out the foregrounds and, therefore, do not seem to represent a major obstacle for the measurement of the global sky signal. This seems also to be confirmed by \cite{voytek14} who achieve an average $\sim$500~mK residual spectra in the 60-90~MHz band after foreground subtraction without including any ionospheric modeling.\\
 
We thank the referee for useful comments that improved the manuscript. GB is grateful to Ravi Subrahmanyan for useful discussions that initiated this work and to Oleg Smirnov for useful discussions on calibration. MM acknowledges support by the National Aeronautics and Space Administration through the Hubble Postdoctoral Fellowship and also from NSF grant AST~1312724. LEDA is supported by NSF grants AST-1106045, AST-1105949, AST-1106059, and AST-1106054.


\begin{thebibliography}		

\bibitem[Atoyan et al.(1995)]{atoyan95} Atoyan, A.~M., Aharonian, F.~A., {V{\"o}lk}, H.~J.\ 1995, \prd, 52, 3265 

\bibitem[\protect\citeauthoryear{{Bernardi} et~al.}{{Bernardi} et~al.}{2013}]{bernardi13}{Bernardi} G. et al., 2013, \apj, 711, 105

\bibitem[\protect\citeauthoryear{{Bernardi} et~al.}{{Bernardi} et~al.}{2011}]{bernardi11}{Bernardi} G., {Mitchell} D.A., {Ord} S.M., {Greenhill} L.J., {Pindor} B., {Wayth} R.B. \& {Wyithe} S.B., 2011, \mnras, 413, 411

\bibitem[\protect\citeauthoryear{{Bernardi} et~al.}{{Bernardi} et~al.}{2010}]{bernardi10}{Bernardi} G. et al., 2010, \aap, 522, 67

\bibitem[\protect\citeauthoryear{{Bernardi} et~al.}{{Bernardi} et~al.}{2009}]{bernardi09}{Bernardi} G. et al., 2009, \aap, 500, 965

\bibitem[\protect\citeauthoryear{{Bowman}, {Hewitt} \& {Rogers}}{{Bowman}, {Hewitt} \& {Rogers}}{2008}]{bowman08}{Bowman} J.D., {Hewitt} J.N. \& {Rogers} A.E.E., 2008, \aj, 676, 1

\bibitem[\protect\citeauthoryear{{Bowman} \& {Rogers}}{{Bowman} \& {Rogers}}{2010}]{bowman10}{Bowman} J.D. \& {Rogers} A.E.E., 2010, \nat, 468, 796

\bibitem[\protect\citeauthoryear{{Burns} et~al.}{{Burns} et~al.}{2011}]{burns11}{Burns} J.O. et al., 2011, arXiv:1106.5194

\bibitem[\protect\citeauthoryear{{Cohen} et~al.}{{Cohen} et~al.}{2007}]{cohen07}{Cohen} A.S., {Lane} W.M., {Cotton} W.D., {Kassim} N. E., {Lazio} T.J.W., {Perley} R.A., {Condon} J.J., \& {Erickson} W.C. 2007, \aj, 134, 1245

\bibitem[\protect\citeauthoryear{{Datta}, {Bowman} \& {Carilli}}{{Datta}, {Bowman} \& {Carilli}}{2010}]{datta10}{Datta} A., {Bowman} J.D. \& {Carilli} C.L., \apj, 724, 526

\bibitem[\protect\citeauthoryear{{de Oliveira-Costa} et al.}{{de Oliveira-Costa} et al.}{2008}]{deoliveiracosta08}{de Oliveira-Costa} A., {Tegmark} M., {Gaensler} B.M., {Jonas} J., {Landecker} T.L. \& {Reich} P., \mnras, 388, 247

\bibitem[\protect\citeauthoryear{{Ellingson}}{{Ellingson}}{2013}]{ellingson13}{Ellingson} S.W., {Craig} J., {Dowell}, J., {Taylor}, G.B., \& Helmboldt, J.F.  2013, IEEE Int'l Symp. on Phased Array Systems and Technology, Boston MA, Oct 2013.

\bibitem[\protect\citeauthoryear{{Ellingson}}{{Ellingson}}{2011}]{ellingson11}{Ellingson} S., 2011, IEEE Trans. Antennas and Propagation, 59, 1855

\bibitem[\protect\citeauthoryear{{Ellingson}}{{Ellingson}}{2010}]{ellingson10}{Ellingson} S., 2010, LWA memo 175, 

\bibitem[\protect\citeauthoryear{{Eisenstein}, {Hu} \& {Tegmark}}{{Eisenstein}, {Hu} \& {Tegmark}}{1998}]{eisenstein99}{Eisenstein} D.J.,  {Hu} W. \& {Tegmark} M., 1999, \apj 518, 2

\bibitem[Di Matteo et al.(2002)]{dimatteo02} Di Matteo, T., Perna, R., Abel, T., \& Rees, M.~J.\ 2002, \apj, 564, 576 

\bibitem[\protect\citeauthoryear{{Dowell}}{{Dowell}}{2011}]{dowell11}{Dowell} J., 2011, LWA memo 178, 

\bibitem[\protect\citeauthoryear{{Fialkov}, {Barkana} \& {Visbal}}{{Fialkov}, {Barkana} \& {Visbal}}{2014}]{fialkov14}{Fialkov} A., {Barkana}, R. \& {Visbal} E., 2014, Nature, 506, 197

\bibitem[\protect\citeauthoryear{{Furlanetto}}{{Furlanetto}}{2006}]{furlanetto06a}{Furlanetto} S.R., 2006, \mnras, 371, 867

\bibitem[\protect\citeauthoryear{{Furlanetto}, {Oh} \& {Briggs}}{{Furlanetto}, {Oh} \& {Briggs}}{2006}]{furlanetto06}{Furlanetto} S.R., {Oh} P.S. \& {Briggs} F.H., 2006, PhR, 433, 181

\bibitem[\protect\citeauthoryear{{Ghosh} et~al.}{{Ghosh} et~al.}{2012}]{ghosh12}{Ghosh} A., {Prasad} J., {Bharadwaj} S., {Ali} S.S. \& {Chengalur} J., 2012, \aj, 426, 3295

\bibitem[\protect\citeauthoryear{{Gnedin} \& {Shaver}}{{Gnedin} \& {Shaver}}{2004}]{gnedin04}{Gnedin} N,Y. \& {Shaver} P.A., 2004, \apj 608, 611

\bibitem[\protect\citeauthoryear{{Greenhill} \& {Bernardi}}{{Greenhill} \& {Bernardi}}{2012}]{greenhill12}{Greenhill} L.J. \& {Bernardi} G.,  2012, 11th Asian-Pacific Regional IAU Meeting 2011, NARIT Conference Series, Vol. 1 eds. S. Komonjinda, Y. Kovalev, and D. Ruffolo (2012), arXiv:1201.1700 

\bibitem[\protect\citeauthoryear{{Hales}, {Baldwin} \& {Warner}}{{Hales}, {Baldwin} \& {Warner}}{1988}]{hales88} {Hales} S.E.G., {Baldwin} J.E. \& {Warner}, 1988, \mnras, 234, 919

\bibitem[\protect\citeauthoryear{{Harker} et~al.}{{Harker} et~al.}{2012}]{harker12}{Harker} G.J.A., {Pritchard} J.R., {Burns} J.O. \& {Bowman} J.D., 2012, \mnras, 419, 1070

\bibitem[\protect\citeauthoryear{{Haslam} et~al.}{{Haslam} et~al.}{1982}]{haslam82}{Haslam} C.G.T., {Salter} C.J., {Stoffel} H. \& {Wilson} W.E., 1982, \aap, 47, 1

\bibitem[\protect\citeauthoryear{{Helmboldt} et~al.}{{Helmboldt} et~al.}{2008}]{helmboldt08}{Helmboldt} J., {Kassim} N., {Cohen} A., {Lane} W. \& {Lazio} T.J. 2008, \apj, 174, 313

\bibitem[\protect\citeauthoryear{{Jelic} et~al.}{{Jelic} et~al.}{2008}]{jelic08}{Jelic} V. et al, 2008, \mnras, 389, 1319

\bibitem[\protect\citeauthoryear{{Kraus}}{{Kraus}}{1950}]{kraus50}{Kraus} J.D., ``Antennas''. New York, McGraw-Hill, 1950

\bibitem[\protect\citeauthoryear{{Landecker} \& {Wielebinski}}{{Landecker} \& {Wielebinski}}{1970}]{landecker70}{Landecker} T.L. \& {Wielebinski} R., 1970, Austrialian Journal of Physics Supplement, 16, 1

\bibitem[\protect\citeauthoryear{{Liu} et~al.}{{Liu} et~al.}{2012}]{liu12}{Liu} A., {Pritchard} J.R., {Tegmark} M.  \& {Loeb} A., 2011, 87, 3002

\bibitem[Madau et al.(1997)]{madau97} Madau, P., Meiksin, A. \& Rees, M.~J.\ 1997, \apj, 475, 429 

\bibitem[Mack \& Wesley(2008)]{mack08} Mack, K.~J. \& Wesley, D.~H.\ 2008, arXiv:0805.1531 

\bibitem[\protect\citeauthoryear{{Mauch} et~al.}{{Mauch} et~al.}{2003}]{mauch03}{Mauch} T., {Murhpy} T., {Buttery} H.J., {Curran} J., {Hunstead} R.W., {Piestrzynski} B., {Robertson} J.G. \& {Sadler} E.M., 2003, \mnras, 342, 1117

\bibitem[\protect\citeauthoryear{{McQuinn} et~al.}{{McQuinn} et~al.}{2006}]{mcquinn06}{McQuinn} M., {Zahn} O., {Zaldarriaga} M., {Hernquist} L. \& {Furlanetto} S.R., 2006, \apj, 653, 815

\bibitem[McQuinn \& O'Leary(2012)]{mcquinn12} McQuinn, M., \& O'Leary, R.~M.\ 2012, \apj, 760, 3 

\bibitem[\protect\citeauthoryear{{Mirocha}}{{Mirocha}}{2014}]{mirocha14}{Mirocha} J., 2014, \mnras, 443, 1211

\bibitem[\protect\citeauthoryear{{Mirocha}, {Harker} \& {Burns}}{{Mirocha}, {Harker} \& {Burns}}{2013}]{mirocha13}{Mirocha} J., {Harker}, G.J.A. \& {Burns}, J.O., 2013, \apj, 777, 118

\bibitem[\protect\citeauthoryear{{Mitchell} et~al.}{{Mitchell} et~al.}{2008}]{mitchell08}{Mitchell} D.A. et al., 2008, IEEE Journal of Selected Topics in Signal Processing, 2, 707

\bibitem[\protect\citeauthoryear{{Morandi} \& {Barkana}}{{Morandi} \& {Barkana}}{2012}]{morandi12}{Morandi} A. \& {Barkana} R., 2012, ApJ, 425, 2551

\bibitem[\protect\citeauthoryear{{Morales} \& {Wyithe}}{{Morales} \& {Wyithe}}{2010}]{morales10}{Morales} M.F. \& {Wyithe} J.S.B., 2010, ARA\&A, 48, 127

\bibitem[\protect\citeauthoryear{{Morales} et~al.}{{Morales} et~al.}{2012}]{morales12}{Morales} M.F., {Hazelton} B., {Sullivan} I. \& {Beardsley} A., 2012, \apj, 752, 137

\bibitem[\protect\citeauthoryear{{Paciga} et~al.}{{Paciga} et~al.}{2013}]{paciga13}{Paciga} G. et al., 2013, \mnras, 433, 639

\bibitem[\protect\citeauthoryear{{Parsons} et~al.}{{Parsons} et~al.}{2012}]{parsons12}{Parsons} A., {Pober} J., {Aguirre} J.E., {Carilli} C.L., {Jacobs} D.C. \& {Moore} D.F., 2012, \apj, 756, 165

\bibitem[\protect\citeauthoryear{{Patra} et~al.}{{Patra} et~al.}{2012}]{patra12}{Patra} N., {Subrahmanyan} R., {Raghunathan} A. \& {Udaya Shankar} N., 2012, ExA, 36, 319

\bibitem[\protect\citeauthoryear{{Petrovic} \& {Oh}}{{Petrovic} \& {Oh}}{2011}]{petrovic11}{Petrovic} N. \& Oh S.P., 2011, \mnras, 413, 2103

\bibitem[\protect\citeauthoryear{{Pober} et~al.}{{Pober} et~al.}{2013}]{pober13}{Pober} J.C. et al., 2013, \apj, 768, 36

\bibitem[\protect\citeauthoryear{{Pritchard} \& {Loeb}}{{Pritchard} \& {Loeb}}{2008}]{pritchard08}{Pritchard} J.R. \& {Loeb} A., 2008, \prd, 78, 3511

\bibitem[\protect\citeauthoryear{{Pritchard} \& {Loeb}}{{Pritchard} \& {Loeb}}{2010}]{pritchard10}{Pritchard} J.R. \& {Loeb} A., 2010, \prd, 82, 3006

\bibitem[\protect\citeauthoryear{{Reich} \& {Reich}}{{Reich} \& {Reich}}{1988}]{reich88}{Reich} W. \& {Reich} P., 1988, \aa, 196, 211

\bibitem[\protect\citeauthoryear{{Rogers} \& {Bowman}}{{Rogers} \& {Bowman}}{2008}]{rogers08}{Rogers} A.E.E. \& {Bowman} J.D., 2008, \aj, 136, 614

\bibitem[\protect\citeauthoryear{{Rogers} \& {Bowman}}{{Rogers} \& {Bowman}}{2012}]{rogers12}{Rogers} A.E.E. \& {Bowman} J.D., 2012, Radio Science, 47, RS0K06

\bibitem[\protect\citeauthoryear{{Rybicki} \& {Lightman}}{{Rybicki} \& {Lightman}}{1979}]{rybicki79}{Rybicki} G.~B., \& {Lightman} A.~P. 1979, New York, Wiley-Interscience, 1979.~393 p.

\bibitem[\protect\citeauthoryear{{Shaver} et~al.}{{Shaver} et~al.}{1999}]{shaver99}{Shaver} P.A., {Windhorst} R.A., {Madau} P. \& {de Bruyn} A.G., 1999, \aap, 380, 390

\bibitem[\protect\citeauthoryear{{Santos}, {Cooray} \& {Knox}}{{Santos}, {Cooray} \& {Knox}}{2005}]{santos05}{Santos} M.G., {Cooray} A. \& {Knox} L., 2005, \apj, 625, 575

\bibitem[\protect\citeauthoryear{{Sethi}}{{Sethi}}{2005}]{sethi05}{Sethi} S.K., 2005, \mnras, 363, 818

\bibitem[\protect\citeauthoryear{{Stanimirovic}}{{Stanimirovic}}{2002}]{stanimirovic02}{Stanimirovic} S., 2002, ASP Conference Proceedings, 278, 375

\bibitem[Strong et al.(2000)]{strong00} Strong, A.~W., Moskalenko, I.~V.\& Reimer, O.\ 2000, \apj, 537, 763 

\bibitem[Strong et al.(2007)]{strong07} Strong, A.~W., Moskalenko, I.~V. \& Ptuskin, V.~S.\ 2007, Annual Review of Nuclear and Particle Science, 57, 285 

\bibitem[\protect\citeauthoryear{{Switzer} \& {Liu}}{{Switzer} \& {Liu}}{2014}]{switzer14}{Switzer} E.R. \& Liu A., 2014, \apj, 793, 102

\bibitem[\protect\citeauthoryear{{Tasse} et~al.}{{Tasse} et~al.}{2013}]{tasse13}{Tasse} C., {van der Tol} S., {van Zwieten} J., {van Diepen} G. \& {Bhatnagar} S., 2013, \aap, 553, 105

\bibitem[\protect\citeauthoryear{{Taylor} et~al.}{{Taylor} et~al.}{2012}]{taylor12}{Taylor} G.B. et al., 2012, JAI, 150004

\bibitem[\protect\citeauthoryear{{Tegmark}}{{Tegmark}}{1997}]{tegmark97}{Tegmark} M., 1997, \prd, 55, 5895

\bibitem[\protect\citeauthoryear{{Tegmark} et~al.}{{Tegmark} et~al.}{2000}]{tegmark00}{Tegmark} M., {Eisenstein}, D. J., {Hu}, W., \& {de Oliveira-Costa}, A. 2000, \apj, 530, 133

\bibitem[\protect\citeauthoryear{{Thompson}, {Moran} \& {Swenson}}{{Thompson}, {Moran} \& {Swenson}}{1990}]{thompson90}{Thompson} A.R., {Moran} J.M. \& {Swenson} G.W., 2001, John Wiley \& Sons 

\bibitem[\protect\citeauthoryear{{Thyagarajan} et~al.}{{Thyagarajan} et~al.}{2013}]{thyagarajan13}{Thyagarajan} N. et al., 2013, \apj, 776, 6 

\bibitem[Trott et al.(2012)]{trott12} Trott C.~M., Wayth R.~B., \& Tingay, S.~J.\ 2012, \apj, 757, 101 

\bibitem[\protect\citeauthoryear{{Valdes} et al.}{{Valdes} et al.}{2007}]{valdes07}{Valdes} M., {Ferrara} A., {Mapelli} M. \& {Ripamonti} E., 2007, \mnras, 377, 245

\bibitem[\protect\citeauthoryear{{Valdes} \& {Ferrara}}{{Valdes} \& {Ferrara}}{2008}]{valdes08}{Valdes} M. \& {Ferrara} A., 2008, \mnras, 387, 8

\bibitem[\protect\citeauthoryear{{Valdes}, {Evoli} \& {Ferrara}}{{Valdes}, {Evoli} \& {Ferrara}}{2010}]{valdes10}{Valdes} M., {Evoli} C. \& {Ferrara} A., 2010, \mnras, 404, 1569

\bibitem[\protect\citeauthoryear{{Valdes} et al.}{{Valdes} et al.}{2013}]{valdes13}{Valdes} M., {Evoli} C., {Mesinger} A., {Ferrara} A. \& {Yoshida} N., 2013, \mnras, 429, 1705

\bibitem[\protect\citeauthoryear{{Vedantham} et al.}{{Vedantham} et al.}{2012}]{vedantham12}{Vedantham} H.K., {Udaya Shankar} N. \& {Subrahmanyan} R., 2012, \apj, 745, 176

\bibitem[\protect\citeauthoryear{{Vedantham} et al.}{{Vedantham} et al.}{2013}]{vedantham13}{Vedantham} H.K., {Koopmans} L.V.E., {de Bruyn} A.G., {Wijnholds} S.J., {Ciardi} B. \& {Brentjens} M.A., 2014, MNRAS, 437, 1056

\bibitem[\protect\citeauthoryear{{Voytek} et al.}{{Voytek} et al.}{2014}]{voytek14}{Voytek} T.C., {Natarajan} A., {Jauregui} G., {Jose} M., {Peterson} J.B. \& {Lopez-Crus} O., 2014, \apj, 782, 9

\bibitem[\protect\citeauthoryear{{Williams} et al.}{{Williams} et al.}{2012}]{williams12}{Williams} C.L. et al., 2012, \apj, 755, 47

\bibitem[\protect\citeauthoryear{{Zaldarriaga}, {Furlanetto} \& {Hernquist}}{{Zaldarriaga}, {Furlanetto} \& {Hernquist}}{2004}]{zaldarriaga04}{Zaldarriaga} M., {Furlanetto} S.R. \& {Hernquist} L., \apjs, 608, 622

\end{thebibliography}
\end{document}